\documentclass[12pt, draftclsnofoot, onecolumn]{IEEEtran}
\usepackage[noadjust]{cite}
\usepackage{amsmath}
\usepackage{algorithmic}
\usepackage{array}
\usepackage{mdwmath}
\usepackage{eqparbox}
\usepackage{amssymb}
\usepackage{stfloats}
\usepackage{mathtools}
\usepackage[T1]{fontenc}
\usepackage[utf8]{inputenc}
\usepackage[font=small]{caption}
\usepackage{url}
\usepackage[pdftex]{graphicx}
\graphicspath{{../pdf/}{../jpeg/}}
\DeclareGraphicsExtensions{.pdf,.jpeg,.png}
\usepackage{epstopdf}
\usepackage{amsfonts}
\usepackage[tight,footnotesize]{subfigure}

\begin{document}

\title{Asymptotic Analysis of RZF over Double Scattering Channels with MMSE Estimation}
\author{Qurrat-Ul-Ain~Nadeem,~\IEEEmembership{Student Member,~IEEE,} Abla~Kammoun,~\IEEEmembership{Member,~IEEE,}
        M{\'e}rouane~Debbah,~\IEEEmembership{Fellow,~IEEE,} and ~ Mohamed-Slim~Alouini,~\IEEEmembership{Fellow,~IEEE}

\thanks{Q.-U.-A. Nadeem, A. Kammoun and M.-S. Alouini are with the Computer, Electrical and Mathematical Sciences and Engineering (CEMSE) Division, Building 1, Level 3, King Abdullah University of Science and Technology (KAUST), Thuwal, Makkah Province, Saudi Arabia 23955-6900 (e-mail: \{qurratulain.nadeem,abla.kammoun,slim.alouini\}@kaust.edu.sa)}
\thanks{M. Debbah is  with  Sup{\'e}lec, Gif-sur-Yvette, France and Mathematical and Algorithmic Sciences Lab, Huawei France R\&D, Paris, France (e-mail: merouane.debbah@huawei.com, merouane.debbah@supelec.fr).}
\thanks{A part of this paper is submitted to IEEE Global Communications Conference (GLOBECOM), Abu Dhabi, UAE, Dec. 2018.}
}

\markboth{}%
{Shell \MakeLowercase{\textit{et al.}}: Bare Demo of IEEEtran.cls for Journals}
\maketitle
\vspace{-.75in}
\begin{abstract}
 This paper studies the ergodic rate performance of regularized zero-forcing (RZF) precoding in the downlink  of a multi-user multiple-input single-output (MISO) system, where the channel between the base station (BS) and each user  is modeled by the double scattering model. This non-Gaussian channel model is a function of both the antenna correlation and the structure of scattering in the propagation environment. This paper makes the preliminary contribution of deriving the minimum-mean-square-error (MMSE) channel estimate for this model. Then under the assumption that the users are divided into groups of common correlation matrices, this paper derives deterministic approximations of the signal-to-interference-plus-noise ratio (SINR) and the ergodic rate, which are almost surely tight in the limit that the number of BS antennas, the number of users and the number of scatterers in each group grow infinitely large. The derived results are expressed in a closed-form for the special case of multi-keyhole channels. Simulation results confirm the close match provided by the asymptotic analysis for moderate system dimensions.  We show that the maximum number of users that can be supported simultaneously, while realizing large-scale MIMO gains, is equal to the number of scatterers. 
\end{abstract}


\vspace{-.2in}
\section{Introduction}

Large-scale multiple-input multiple-output (MIMO) is widely considered as a promising technology for next generation wireless communication systems \cite{massive2, massiveMIMOO, massive_Luca, massive_Luca1, massive_Luca2, massiveMIMObook1,SINRdeterministic,linearprecoding,massiveMIMObook}. However, most works on this subject share the underlying assumption of rich scattering conditions and, thus, work with full rank Rayleigh or Rician fading channel matrices. Although the use of full rank channel matrices facilitates the derivation of  closed-form capacity bounds and approximations, these models do not capture the characteristics of realistic propagation environments, where the presence of spatial correlation and poor scattering conditions significantly affects the system performance \cite{gesbert, gesbert1, correlationpresence3}.

Many works have already considered correlated Rayleigh fading channel models in their analysis of large-scale and Massive MIMO systems \cite{massiveMIMOO, SINRdeterministic,  massive_Luca, massive_Luca1, massive_Luca2, linearprecoding}.  The authors in \cite{massiveMIMOO} show that in the limit of an infinite number of antennas, the effects of noise, channel estimation errors, and interference vanish, while pilot contamination remains the only performance limitation. However, very recently the authors in \cite{massive_Luca} show in the same limit that the capacity increases without bound even under pilot contamination, provided that the spatial correlation matrices of the pilot contaminating users are asymptotically linearly independent and multi-cell minimum-mean-square-error (MMSE) precoding/combining techniques are employed. 


Despite the important role played by spatial correlation in determining large-scale MIMO performance, low rank channels have been observed in MIMO systems that have low antenna correlation at both ends of the transmission link \cite{scattering}. In fact, Gesbert \textit{et al.} show in \cite{gesbert} that MIMO capacity is governed by both the spatial correlation at the communication ends and the structure of scattering in the propagation environment. Motivated by this, the authors devise a ``double scattering channel model'', which utilizes the geometry of the propagation environment to model spatial correlation, rank deficiency and limited scattering.  Unlike the commonly utilized MIMO channel models, the double scattering model is non-Gaussian, rendering the theoretical analysis of techniques designed using this model extremely hard. 

\vspace{-.23in}
\subsection{Related Literature}
\vspace{-.04in}
The main literature related to the double scattering channel model is scarce and is mainly represented by \cite{doublescattering2, keyholee2, doublescattering1, doublescattering3, doublescattering4, doublescattering6, scattering, levin, doublescattering}. The authors in \cite{doublescattering2} study its diversity order and show that a MIMO system with $t$ transmit (Tx) and $r$ receive (Rx) antennas and $s$ scatterers achieves a diversity of order $trs/ \max(t, r, s)$.  The authors in \cite{doublescattering1} analyze the ergodic MIMO capacity taking into account the presence of  double scattering and keyhole effects and show that the use of multiple antennas in keyhole channels only offers diversity gains, but no spatial multiplexing gains. Closed-form upper-bounds on the sum-capacity are obtained in \cite{doublescattering6} for the MIMO multiple access channel with double scattering fading.


A few papers have analyzed the double scattering model in large-scale MIMO settings \cite{scattering, doublescattering, ourworkRMT1}. The authors in \cite{scattering} study this model without Tx and Rx correlation using tools from free probability theory and derive implicit expressions for the asymptotic mutual information (MI) and signal-to-interference-plus-noise ratio (SINR) of the MMSE detector. The authors in \cite{doublescattering} study a MIMO multiple access system with double-scattering channels and use random matrix theory (RMT) tools to derive an almost surely tight deterministic approximation of the MI. However, none of these papers consider practical large-scale multi-user  MIMO systems with linear signal processing schemes and channel estimation.  The authors are only aware of \cite{doublescattering_multicell}, that deals with these aspects using numerical examples instead of theoretical analysis.


\vspace{-.15in}
\subsection{Main Contributions}

The focus of this work is on the downlink (DL) of a single-cell multi-user multiple-input single-output (MISO) system. We consider double scattering fading between the BS and the users in its most general form.  The users are further divided into $G$ groups, such that the users in the same group are characterized by common correlation matrices. This assumption is needed to provide some level of analytical tractability for an, otherwise, extremely difficult to study non-Gaussian channel model. We stress here that having variation across the correlation matrices is important in realizing the fundamental performance of large-scale MIMO systems as shown in \cite{massive_Luca, massive_Luca1, massiveMIMObook}. However, these works base their analysis on the correlated Rayleigh channel model, which encourages us to study the benefits of having a large number of antennas under the more realistic double scattering model, where we focus on studying the effects of the scattering conditions (instead of the structure of the correlation matrices) on the system performance. 

We consider a realistic large-scale MIMO system where the BS obtains channel state information (CSI) from uplink pilot transmissions and applies MMSE estimation technique. Under the assumption that the number of BS antennas, scatterers and users grow large, we derive asymptotically tight deterministic approximations of the SINR and the ergodic rate with  regularized zero-forcing (RZF) precoding, which are accurate for moderate system dimensions as well, as shown through simulations. The deterministic approximations are expressed in a closed-form for the multi-keyhole channel and some important insights into the impact of different system parameters on the sum rate performance are drawn. We show that the number of users scheduled simultaneously should be less than the number of scatterers to see `large-scale MIMO gains'.

\vspace{-.15in}
\subsection{Outline and Notation}

The rest of the paper is organized as follows. Section II presents the transmission model and introduces the double scattering channel model along with its MMSE estimate. In Section III, the asymptotically tight deterministic equivalents of the SINR and user rates with RZF precoding are derived.  Simulation results are provided in Section IV and Section V concludes the paper. 


The following notation is used. Boldface lower-case and upper-case characters denote vectors and matrices respectively.  The operator $\text{tr } (\textbf{X})$ denotes the trace of a matrix \textbf{X}. The spectral norm of a matrix \textbf{X} is denoted by $||\textbf{X}||$ and the $N\times N$ diagonal matrix of entries $\{x_{n}\}$ is denoted by $\textbf{X}=\text{diag}(x_{1}, x_{2},\dots, x_{N})$. A random vector $\textbf{x} \sim \mathcal{CN} (\textbf{m},\boldsymbol{\Phi})$ is complex Gaussian distributed with mean $\textbf{m}$ and covariance matrix $\boldsymbol{\Phi}$. The notation $ \xrightarrow[]{a.s.}$ denotes almost sure convergence.

\vspace{-.12in}
\section{System Model}
Consider a single-cell multi-user MISO system where a BS equipped with $N$ antennas serves $K$ single-antenna users, who are divided into $G$ groups of $K_{g}$, $g=1, \dots, G$, users such that the users in the same group experience similar propagation conditions. In this section, we outline the transmission model, and introduce the double scattering channel along with its MMSE estimate. 

\vspace{-.15in}
\subsection{Transmission Model}

The received complex baseband signal $y_{k,g}$ at user $k$ in group $g$ is given as,
\begin{align}
\label{Rx_sig}
&y_{k,g}=\textbf{h}_{k,g}^{H}\textbf{x} + {n}_{k,g}, \hspace{.1in} k=1, \dots, K_{g}, \hspace{.06in} g=1, \dots, G,
\end{align}
where $\textbf{h}_{k,g}^{H} \in \mathbb{C}^{1 \times N}$ is the channel vector from the BS to user $k$ in group $g$, $\textbf{x} \in \mathbb{C}^{N\times 1}$ is the Tx signal vector and $n_{k,g} \sim \mathcal{CN}(0,\sigma^{2})$ is the receiver noise. The Tx signal vector \textbf{x} is given as,
\begin{align}
&\textbf{x}=\sum_{g=1}^{G} \sum_{k=1}^{K_{g}} \sqrt{p_{k,g}} \textbf{g}_{k,g} s_{k,g},
\end{align}
where $\textbf{g}_{k,g} \in \mathbb{C}^{N\times 1}$, $p_{k,g} \geq 0$ and $s_{k,g} \sim \mathcal{CN}(0,1)$ are the precoding vector, signal power  and  data symbol for user $k$ in group $g$ respectively. The precoding vectors satisfy,
\begin{align}
\label{p_cons}
& \mathbb{E}[||\textbf{x}||^{2}]=\mathbb{E}[\text{tr }(\textbf{P} \textbf{G}^{H} \textbf{G})] \leq \bar{P},
\end{align}
where $\bar{P}>0$ is the average total Tx power, $\textbf{P}=\text{diag}(p_{1,1},  \dots, p_{K_{1},1}, p_{1,2}, \dots, p_{K_{G},G})$ $\in \mathbb{R}^{K\times K}$ and $\textbf{G}=[\textbf{G}_{1},  \dots, \textbf{G}_{G}]  \in \mathbb{C}^{N\times K}$ is the precoding matrix, where $\textbf{G}_{g} = [\textbf{g}_{1,g}, \dots, \textbf{g}_{K_{g},g}]$ $\in \mathbb{C}^{N\times K_{g}}$.

\begin{figure}[!t]
\centering
\includegraphics[width=2.5 in]{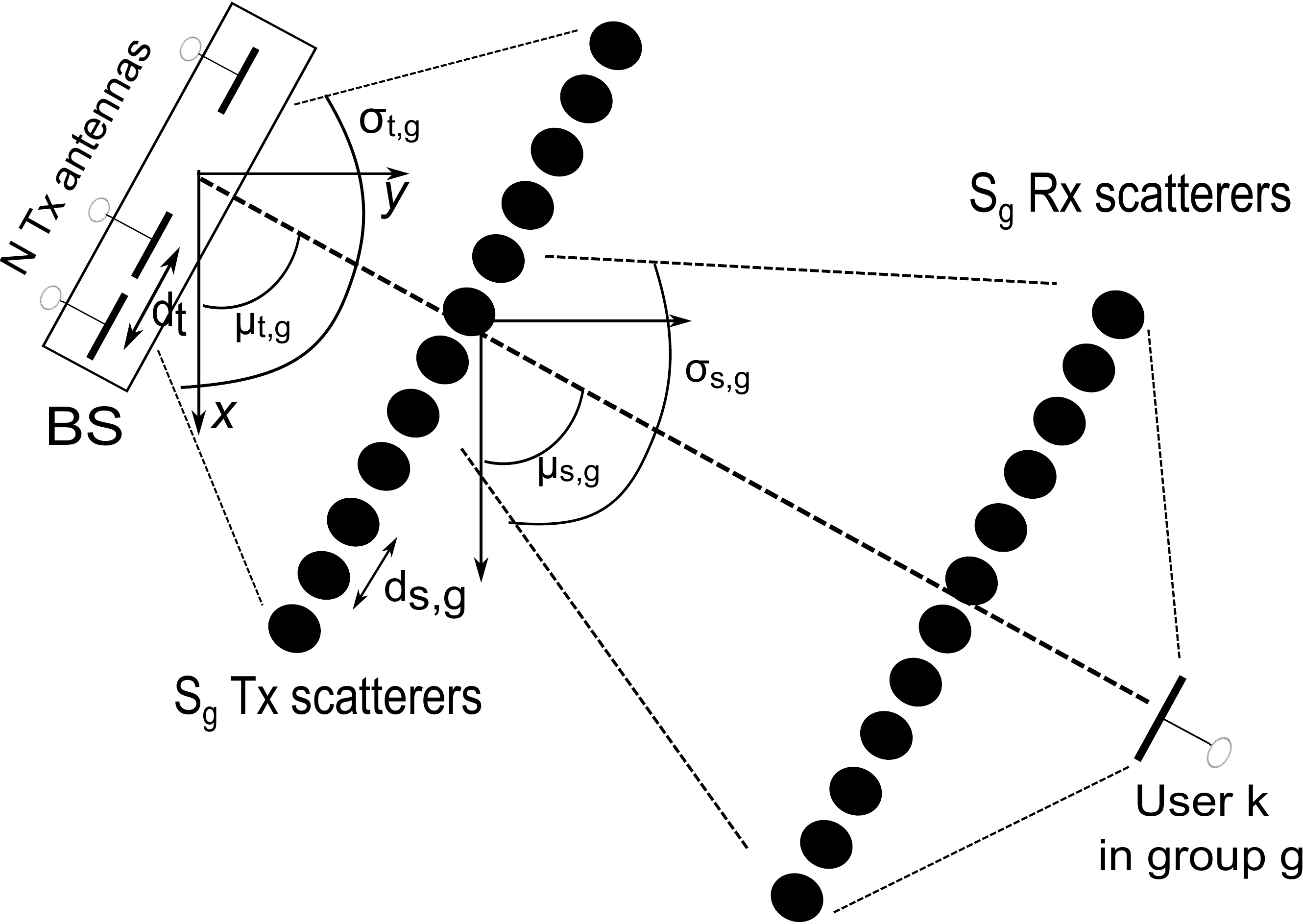}
\caption{Geometric model of the double scattering channel between the BS and the user $k$ in group $g$.}
\label{scattering_model}
\end{figure}

\vspace{-.2in}
\subsection{Double Scattering Channel Model}
A main contribution of this paper is to apply the double scattering channel model proposed in \cite{gesbert} to a multi-user MISO system. This non-Gaussian model has rank that is determined by both the spatial correlation between the antennas at the BS and the structure of scattering in the propagation environment.  The double scattering channel vector $\textbf{h}_{k,g}$ is given as \cite{gesbert},
\begin{align}
\label{DS}
& \textbf{h}_{k,g}=\sqrt{S_{g}} \left(\frac{1}{\sqrt{S_{g}}} \textbf{R}_{BS_{g}}^{1/2} \textbf{W}_{g} \bar{\textbf{S}}_{g}^{1/2}  \right) \tilde{\textbf{w}}_{k,g}, 
\end{align}
where $S_{g}$ is the number of scatterers at the Tx and the Rx sides in group $g$, $\textbf{R}_{BS_{g}} \in \mathbb{C}^{N \times N}$ is the correlation matrix between the BS antennas and the $S_{g}$ Tx scatterers, $\bar{\textbf{S}}_{g} \in \mathbb{C}^{S_{g}\times S_{g}}$ is the correlation matrix between the $S_{g}$ Tx and Rx scatterers, $\textbf{W}_{g} \in \mathbb{C}^{N\times S_{g}}$ is a standard complex Gaussian matrix that describes the small-scale fading between the BS and the scattering cluster at the Tx side  and $\tilde{\textbf{w}}_{k,g} \sim  \mathcal{CN}(\boldsymbol{0},\frac{1}{S_{g}}\textbf{I}_{S_{g}}) \in \mathbb{C}^{S_{g}\times 1}$ describes the small-scale fading between the user $k$ in group $g$ and the scattering cluster at the Rx side. 
Note that we can assume $\bar{\textbf{S}}_{g}$ to be diagonal without  any loss of generality for the statistics of the received signal. 

Schematic of the double scattering channel is shown in Fig. \ref{scattering_model}, where $\sigma_{t,g}$ and $\sigma_{s,g}$ represent the angular spread of the radiated signal from the BS array and the Tx scatterers respectively in group $g$, and $\mu_{t,g}$ and $\mu_{s,g}$ represent the mean angle of departure (AoD) of the radiated signal from the BS array and the Tx scatterers respectively in group $g$, where $\mu_{t,g}=\mu_{s,g}$. 


\vspace{-.15in}

\subsection{Channel Estimation}

During a dedicated uplink training phase, the users transmit mutually orthogonal pilot sequences that allow the BS to compute the MMSE estimates $\hat{\textbf{h}}_{k,g}$ of the channel vectors $\textbf{h}_{k,g}$. After correlating the received training signal with the pilot sequence of user $k$ in group $g$, the BS estimates the channel vector $\textbf{h}_{k,g}$ based on the received observation, $\textbf{y}^{tr}_{k,g} \in \mathbb{C}^{N\times1}$, given as,
\begin{align}
\label{y_tr}
\textbf{y}_{k,g}^{tr}=\textbf{h}_{k,g}+\frac{1}{\sqrt{\rho_{tr}}} \textbf{n}^{tr}_{k,g},
\end{align}
where $\textbf{n}^{tr}_{k,g} \sim \mathcal{CN} (\boldsymbol{0},\textbf{I}_{N})$ and $\rho_{tr}>0$ is the effective training SNR, assumed to be given here.

\textit{Lemma 1: } The MMSE estimate $\hat{\textbf{h}}_{k,g} \in \mathbb{C}^{N\times 1}$ of the channel $\textbf{h}_{k,g}$ in (\ref{DS}) is given as,
\vspace{-.1in}
\begin{align}
\label{MMSE_estimate}
&\hat{\textbf{h}}_{k,g}=d_{g} \textbf{R}_{BS_{g}} \textbf{Q}_{g} \textbf{y}_{k,g}^{tr}, 
\end{align}
where $d_{g}=\frac{1}{S_{g}} (\text{tr } \bar{\textbf{S}}_{g})$ and $\textbf{Q}_{g}=\left(d_{g} \textbf{R}_{BS_{g}} + \frac{1}{\rho_{tr}} \textbf{I}_{N}\right)^{-1}$.

The proof is postponed to Appendix A. 

We stress that Lemma 1 has been derived for a non-Gaussian channel.  Under the orthogonality property of the MMSE estimate, $\textbf{h}_{k,g}=\hat{\textbf{h}}_{k,g}+\tilde{\textbf{h}}_{k,g}$, where $\tilde{\textbf{h}}_{k,g}$ is the uncorrelated channel estimation error. Since our focus is on a single-cell system so the derived channel estimate is not corrupted by pilot contamination from adjacent cells. Therefore the considered system is referred to  as a large-scale MIMO system instead of a Massive MIMO system.


\vspace{-.16in}
\subsection{Achievable Rates}
\vspace{-.05in}
Since the BS does not have CSI, so it utilizes the MMSE estimate in (\ref{MMSE_estimate}) to implement digital precoding. The analysis presented in this paper is, therefore, a function of $\rho_{tr}$, which determines the error that is incurred in channel estimation. However, to compute the SINR in the downlink, the user needs to have CSI too. A well-known feature of large-scale MIMO systems is channel hardening, which means that the effective useful channel $\textbf{h}_{k,g}^{H}\textbf{g}_{k,g}$ of a user converges to its average value when $N,S$ grow large. Hence, it is sufficient for each user to have only the statistical CSI and the resulting performance loss vanishes in the large system limit. Using this idea, an ergodic achievable user rate can be computed using a technique from \cite{Medard}, widely applied to large-scale MIMO systems in \cite{massive2, multicellTDD, massiveMIMOO,linearprecoding}. The main idea is to decompose $y_{k,g}$ in (\ref{Rx_sig}) as $\sqrt{p_{k,g}} \mathbb{E}[\textbf{h}_{k,g}^{H} \textbf{g}_{k,g}] s_{k,g} + \sqrt{p_{k,g}} (\textbf{h}_{k,g}^{H} \textbf{g}_{k,g}-\mathbb{E}[\textbf{h}_{k,g}^{H} \textbf{g}_{k,g}]) s_{k,g} +\sum_{(k',g')\neq (k,g)} \sqrt{p_{k',g'}} \textbf{h}_{k,g}^{H} \textbf{g}_{k',g'} s_{k',g'} +{n}_{k,g}$ and assume that the average effective channel $\mathbb{E}[\textbf{h}_{k,g}^{H} \textbf{g}_{k,g}]$ is perfectly known at the corresponding user. By treating the inter-user interference and channel uncertainty as worst-case Gaussian noise, the user $k$ in group $g$ can achieve the ergodic rate, \vspace{-.07in}
\begin{align}
\label{rate_11}
R_{k,g}=\log(1+\gamma_{k,g}),
\end{align}
without knowing the instantaneous values of $\textbf{h}^{H}_{k,g} \textbf{g}_{k,g}$. The parameter $\gamma_{k,g}$ can be interpreted as the effective average SINR  of user $k$ in group $g$ and is defined as,
\vspace{-.05in}
\begin{align}
\label{SINR}
& \gamma_{k,g}=\frac{p_{k,g} |\mathbb{E}[\textbf{h}_{k,g}^{H} \textbf{g}_{k,g}]|^{2}}{p_{k,g} \text{var}[\textbf{h}_{k,g}^{H} \textbf{g}_{k,g}] + \sum_{(k',g')\neq (k,g)} p_{k',g'} \mathbb{E} [|\textbf{h}_{k,g}^{H} \textbf{g}_{k',g'}|^{2}]+ \sigma^{2}}.
\end{align}
The ergodic achievable sum rate is given as,
\begin{align}
\label{summm}
R_{sum}=\sum_{g=1}^{G}\sum_{k=1}^{K_{g}} R_{k,g}.
\end{align}

This paper considers RZF precoding, which is a state-of-the-art heuristic precoding scheme with a simple closed-form  expression given as \cite{massiveMIMOO, SINRdeterministic},
\begin{align}
\label{RZF}
&\textbf{G}=\zeta (\hat{\textbf{H}}^{H} \hat{\textbf{H}} + K \alpha \textbf{I}_{N})^{-1} \hat{\textbf{H}}^{H},
\end{align}
where $\hat{\textbf{H}}=[\hat{\textbf{H}}_{1}^{H}, \hat{\textbf{H}}_{2}^{H}, \dots, \hat{\textbf{H}}_{G}^{H}]^{H}  \in \mathbb{C}^{K \times N}$, where $\hat{\textbf{H}}_{g}=[\hat{\textbf{h}}_{1,g}, \hat{\textbf{h}}_{2,g},$ $\dots, \hat{\textbf{h}}_{K_{g},g}]^{H} \in \mathbb{C}^{K_{g} \times N}$, $\alpha$ is the regularization parameter and $\zeta$ ensures that the power constraint in (\ref{p_cons}) is satisfied as,
\begin{align}
\label{zeta}
&\zeta^{2}=\bar{P}/\mathbb{E}[\text{tr} (\textbf{P} \hat{\textbf{H}} (\hat{\textbf{H}}^{H} \hat{\textbf{H}} + K \alpha \textbf{I}_{N})^{-2} \hat{\textbf{H}}^{H})]= \bar{P}/\Theta,
\end{align}
where $\Theta=\mathbb{E}[\text{tr} (\textbf{P} \hat{\textbf{H}} \hat{\textbf{V}}^{2} \hat{\textbf{H}}^{H})]$, where $\hat{\textbf{V}}=(\hat{\textbf{H}}^{H} \hat{\textbf{H}} + K \alpha \textbf{I}_{N})^{-1}$. The SINR in (\ref{SINR}) is  now defined as,
\begin{align}
\label{SINR_RZF}
&\gamma_{k,g\text{RZF}}=\frac{p_{k,g} |\mathbb{E}[\textbf{h}_{k,g}^{H} \hat{\textbf{V}} \hat{\textbf{h}}_{k,g}]|^{2}}{\mathbb{E}[\textbf{h}_{k,g}^{H} \hat{\textbf{V}} \hat{\textbf{H}}_{[k,g]}^{H} \textbf{P}_{[k,g]} \hat{\textbf{H}}_{[k,g]} \hat{\textbf{V}} \textbf{h}_{k,g}] + p_{k,g} \text{var }[\textbf{h}_{k,g}^{H} \hat{\textbf{V}} \hat{\textbf{h}}_{k,g}] + \frac{\Theta}{\rho}},
\end{align}
where $\rho=\frac{\bar{P}}{\sigma^{2}}$, $\hat{\textbf{H}}_{[k,g]}=[\hat{\textbf{H}}_{1}^{H}, \dots, \hat{\textbf{H}}_{g-1}^{H}, \hat{\textbf{h}}_{1,g}, \dots, \hat{\textbf{h}}_{k-1,g}, \hat{\textbf{h}}_{k+1,g},\dots, \hat{\textbf{h}}_{K_{g},g}, \dots, \hat{\textbf{H}}_{G}^{H}]^{H} \in \mathbb{C}^{K-1\times N}$ and $\textbf{P}_{[k,g]}=\text{diag}(p_{1,1}, \dots, p_{K_{g-1},g-1}, p_{1,g}, \dots, p_{k-1,g}, p_{k+1,g},\dots, p_{K_{g},g}, \dots, p_{K_{G},G}) \in \mathbb{C}^{K-1, K-1}$.

\vspace{-.1in}
\section{Main Results}
\vspace{-.05in}
The performance of the users are characterized by their ergodic rates, $R_{k,g}$, which depend on their SINRs. These ergodic rates, with RZF precoding, are difficult to compute for finite system dimensions. However, they tend to deterministic quantities in the large ($N,K$) regime, as shown in \cite{massiveMIMOO, linearprecoding, SINRdeterministic}. These quantities depend only on the statistics of the channels and are referred to as deterministic equivalents. These deterministic equivalents are almost surely (a.s.) tight in the asymptotic limit. By a.s. tight in the asymptotic limit we mean that as the system dimensions grow to infinity, the approximations yielded by these deterministic equivalents tend to the actual values with probability one. Deterministic equivalents were first proposed by Hachem et al. in \cite{walid_mutual_information}, who showed their ability to capture important system performance indicators.


In the sequel, the asymptotic limit, denoted as $N \rightarrow \infty$, represents the following assumption.

\textbf{A-1.} For all $g$, $N$, $S_{g}$, $K_{g}$ and $K$ tend to infinity such that, \vspace{-.05in}
\begin{align}
& 0 < \text{lim inf } \frac{S_{g}}{N} \leq \text{lim sup } \frac{S_{g}}{N} < \infty, \hspace{.05in}  0 < \text{lim inf } \frac{K_{g}}{N} \leq \text{lim sup } \frac{K_{g}}{N} < \infty. \nonumber
\end{align} 
Also, in the sequel, the deterministic equivalent of a sequence of RVs $X_{N}$ is represented by the deterministic sequence $X^{o}_{N}$, which approximates $X_{N}$ such that, \vspace{-.1in}
\begin{align}
&X_{N}-X_{N}^{o} \xrightarrow[N \rightarrow \infty]{a.s.} 0.
\end{align} 

The objective of this section is to derive the deterministic equivalent $\gamma^{o}_{k,g\text{RZF}}$ of the SINR $\gamma_{k,g\text{RZF}}$ with RZF precoding, such that, \vspace{-.1in}
\begin{align}
\label{SS}
& \gamma_{k,g\text{RZF}} -\gamma^{o}_{k,g\text{RZF}} \xrightarrow[N \rightarrow \infty]{a.s.} 0.
\end{align} 
Note that $\gamma_{k,g\text{RZF}} $ is a function of both $\tilde{\textbf{w}}_{k,g}$ and $\textbf{W}_{g}$. From now on, these random vectors and matrices must be understood as sequences of vectors and matrices of growing dimensions. For the sake of simplicity, their dependence on $N$, $S$ and $K$ is not explicitly shown.

We stress again that the result in (\ref{SS}) must be understood in the way that, for each given set of system parameters $N$, $S$, and $K$, we provide approximations of the SINR and the user rates that become increasingly tight as $N$, $S$ and $K$ grow large. We will show later by simulations that these approximations are very accurate for moderate system dimensions - an observation made in \cite{massiveMIMOO, SINRdeterministic,  massive_Luca, linearprecoding, walid_mutual_information} as well. The asymptotic analysis requires the following two assumptions.

\textbf{A-2.} For all $g$, $\text{lim sup}_{N} ||\textbf{R}_{BS_{g}}|| < \infty$ and $\text{lim sup}_{S_{g}} ||\bar{\textbf{S}}_{g}|| < \infty$. 

\textbf{A-3.} For all $g$, $\text{lim inf}_{N} \frac{1}{N} \text{tr } \textbf{R}_{BS_{g}} > 0$ and $\text{lim inf}_{S_{g}} \frac{1}{N} \text{tr } \bar{\textbf{S}}_{g} > 0$.

We first present the Fubini theorem, which is a key mathematical idea behind our derivations. 

\vspace{-.18in}
\subsection{Fubini Theorem} 
\vspace{-.05in}
Most works derive deterministic equivalents for random matrix models created from sums of independent random matrices. In many practical cases, like for the double scattering model, it is necessary to consider products of random matrices. To handle such models, we rely on the concept of iterative deterministic equivalents which utilizes the Fubini theorem \cite{iterative, doublescattering}.

A consequence of the Fubini theorem is that if we have a function $f_{N}(\textbf{H}'_{N}, \textbf{H}''_{N})$ of two independent random sequences of matrices, $(\textbf{H}''_{N})_{N\geq 1}$ and $(\textbf{H}'_{N})_{N\geq 1}$, then we can condition on one sequence, let's say $(\textbf{H}'_{N})_{N\geq 1}$, and find the deterministic equivalent $\tilde{g}_{N}(\textbf{H}'_{N})$ for $f_{N}$. If it can be proved that this deterministic equivalent holds true for every $(\textbf{H}'_{N})_{N\geq 1}$ generated by a space $\Omega$, then it is also valid for the random sequence $(\textbf{H}'_{N}, \textbf{H}''_{N})_{N\geq 1}$ \cite{iterative}. However $\tilde{g}_{N}(\textbf{H}'_{N})$ is still random due to dependence on $(\textbf{H}'_{N})_{N\geq 1}$, so we need to get a deterministic equivalent $g_{N}$ for it.

This is the main mathematical idea behind our derivation of the deterministic equivalent of $\gamma_{k,g\text{RZF}}$ which is a function of the random matrices $\textbf{W}_{g}$ and the random vectors $\tilde{\textbf{w}}_{k,g}$. First, we interpret the double-scattering channel model in (\ref{DS}) as,
\begin{align}
\label{DS1}
& \textbf{h}_{k,g}= \sqrt{S_{g}} \textbf{Z}_{g} \tilde{\textbf{w}}_{k,g},
\end{align}
where $\textbf{Z}_{g}=\frac{1}{\sqrt{S_{g}}} \textbf{R}_{BS_{g}}^{1/2} \textbf{W}_{g} \bar{\textbf{S}}_{g}^{1/2}$, and we assume $\textbf{Z}_{g}$ to be deterministic. Under this setting, the estimate of the double scattering model  in (\ref{MMSE_estimate}) can be interpreted as,
\begin{align}
\label{MMSE_estimate1}
& \hat{\textbf{h}}_{k,g}= \boldsymbol{\Phi}^{1/2}_{g} \bar{\textbf{q}}_{k,g},
\end{align}
where $\bar{\textbf{q}}_{k,g}\sim \mathcal{CN}(\boldsymbol{0},\textbf{I}_{N})$ and $\boldsymbol{\Phi}_{g}$ is the covariance matrix of the channel estimate given as,
\begin{align}
\label{MMSE_estimatecorr}
&\boldsymbol{\Phi}_{g}=d_{g}^{2} \textbf{R}_{BS_{g}} \textbf{Q}_{g} \left( \textbf{Z}_{g} \textbf{Z}^{H}_{g} + \frac{1}{\rho_{tr}} \textbf{I}_{N}\right)  \textbf{Q}^{H}_{g} \textbf{R}^{H}_{BS_{g}}.
\end{align}
We obtain the deterministic equivalent of $\gamma_{k,g\text{RZF}}$ in terms of certain fixed point equations that depend on the deterministic matrices $\textbf{Z}_{g}$'s using RMT results from \cite{theorem}. We then extend the analysis by allowing $\textbf{Z}_{g}$'s to be random based on the Fubini theorem. In this second step, we derive the deterministic equivalents of the fixed point equations under the actual random $\textbf{Z}_{g}$'s. 

To summarize, the ``randomness" related to $\textbf{W}_{g}$ is first removed and then introduced later after we have the deterministic equivalent of the SINR as a function of the random vector $\tilde{\textbf{w}}_{k,g}$ and the deterministic matrix $\textbf{W}_{g}$. From this construction, the resulting deterministic approximation is referred to as an iterative deterministic equivalent.

\vspace{-.18in}
\subsection{New and Useful Results} 
\vspace{-.05in}
The two theorems in this section represent the major contributions of this work as they are required to cope with the non-Gaussian channel model in (\ref{DS}) and its estimate in (\ref{MMSE_estimate}) and form the mathematical basis of the subsequent large system analysis of RZF precoding. They provide deterministic equivalents of normalized traces and quadratic forms  involving single and double occurrences of the resolvent matrix $\hat{\textbf{C}}^{-1}(\alpha)$, which is defined as $\hat{\textbf{C}}^{-1}(\alpha)=\left(\frac{1}{K}\hat{\textbf{H}}^{H}\hat{\textbf{H}}+\alpha \textbf{I}_{N}\right)^{-1}$.

The resolvent is of constant use in this paper as it arises from the expression of $\hat{\textbf{V}}$ in the RZF precoder in (\ref{RZF}) and appears in all the terms of the SINR in (\ref{SINR_RZF}). The trace of the resolvent is characterized by the quantities $(m_{g}(\alpha), \bar{m}_{g}(\alpha), \delta_{g}(\textbf{R}_{g}, \alpha))$, $1 \leq g \leq G$  in the asymptotic limit.  The deterministic approximation of the SINR explicitly depends  on these quantities, which is why they are introduced first in the following theorem.

\textit{Theorem 1:} Consider the resolvent matrix, $\hat{\textbf{C}}^{-1}(\alpha)$, where the columns of $\hat{\textbf{H}}^{H}$ are distributed as (\ref{MMSE_estimate}). Then the following system of $3G$ implicit equations in $m_{g}(\alpha), \bar{m}_{g}(\alpha)$ and $\delta_{g}(\textbf{R}_{g}, \alpha)$,
\begin{align}
& m_{g}(\alpha)=\frac{1}{K} d_{g}^2 \left( \sum_{j=1}^{S_{g}} \frac{\bar{s}_{g,j} \delta_{g}(\bar{\textbf{D}}_{g},\alpha)}{1+\frac{K_{g}}{K} d_{g}^{2} \bar{s}_{g,j} \bar{m}_{g}(\alpha) \delta_{g}(\bar{\textbf{D}}_{g},\alpha)} + \frac{S_{g}}{\rho_{tr}} \delta_{g}(\tilde{\textbf{D}}_{g},\alpha)\right), \nonumber \\
\label{fixedpoint}
& \bar{m}_{g}(\alpha)=\frac{1}{1+m_{g}(\alpha)}, \\
& \delta_{g}(\textbf{R}_{g}, \alpha)=\frac{1}{S_{g}} \text{tr } \textbf{R}_{g} \bar{\textbf{T}}(\alpha), \nonumber \\
\label{T_bar}
&\hspace{-.12in}\text{where, } \bar{\textbf{T}}(\alpha)=\left( \sum_{i=1}^{G} \frac{ \bar{\textbf{D}}_{i}}{S_{i}} \left(  \sum_{j=1}^{S_{i}} \frac{ \frac{K_{i}}{K} d_{i}^{2} \bar{s}_{i,j} \bar{m}_{i}(\alpha)}{1+\frac{K_{i}}{K} d_{i}^{2} \bar{s}_{i,j} \bar{m}_{i}(\alpha) \delta_{i}(\bar{\textbf{D}}_{i},\alpha)} \right) + \frac{K_{i}}{K} d_{i}^2 \frac{\bar{m}_{i}(\alpha)}{\rho_{tr}} \tilde{\textbf{D}}_{i} +\alpha \textbf{I}_{N}  \right)^{-1},
\end{align}
has a unique solution satisfying  $(m_{g}(\alpha), \bar{m}_{g}(\alpha)$, $\delta_{g}(\textbf{R}_{g}, \alpha))> 0$ for all $g$ and $\alpha > 0$, where $\textbf{R}_{g}$ is an arbitrary matrix with a uniformly bounded spectral norm, $\bar{\textbf{D}}_{g}=\textbf{R}_{BS_{g}}\textbf{Q}_{g} \textbf{R}_{BS_{g}} \textbf{Q}^{H}_{g} \textbf{R}^{H}_{BS_{g}}$ and $\tilde{\textbf{D}}_{g}=\textbf{R}_{BS_{g}}\textbf{Q}_{g} \textbf{Q}^{H}_{g} \textbf{R}^{H}_{BS_{g}}$.
Let \textbf{U} be any deterministic matrix with a uniformly bounded spectral norm. Under assumptions \textbf{A-1}, \textbf{A-2}, \textbf{A-3} and for $\alpha>0$, we have
\begin{align}
\label{detttt}
\frac{1}{K}\text{tr}(\textbf{U}\hat{\textbf{C}}^{-1}(\alpha))-\frac{1}{K}\text{tr}(\textbf{U}\bar{\textbf{T}}(\alpha))  \xrightarrow[N \rightarrow \infty]{a.s.} 0.
\end{align}

\textit{Proof:} The proof of \textit{Theorem 1} is postponed to Appendix B.

Next we present the deterministic equivalents of $\frac{1}{K}\hat{\textbf{h}}^{H}_{k,g}  \hat{\textbf{C}}_{[k,g]}^{-1} \hat{\textbf{h}}_{k,g}$ and $\frac{1}{K}{\textbf{h}}^{H}_{k,g}  \hat{\textbf{C}}_{[k,g]}^{-1} \hat{\textbf{h}}_{k,g}$, where $\hat{\textbf{C}}_{[k,g]}=\frac{1}{K} \hat{\textbf{H}}_{[k,g]}^{H} \hat{\textbf{H}}_{[k,g]} + \alpha \textbf{I}_{N}$, where $\hat{\textbf{H}}_{[k,g]}$ is defined in Section II-D. These quantities arise in the expression of $\gamma_{k,g\text{RZF}}$ and  appear repeatedly in our analysis. 

\textit{Lemma 2:}  Under the setting of assumptions \textbf{A-1}, \textbf{A-2} and \textbf{A-3} and for $\alpha>0$,
\begin{align}
\label{quad1}
&\frac{1}{K}\hat{\textbf{h}}^{H}_{k,g}  \hat{\textbf{C}}_{[k,g]}^{-1} \hat{\textbf{h}}_{k,g} - m_{g}(\alpha) \xrightarrow[N \rightarrow \infty]{a.s.} 0, \\
\label{quad2}
&\frac{1}{K}{\textbf{h}}^{H}_{k,g}  \hat{\textbf{C}}_{[k,g]}^{-1} \hat{\textbf{h}}_{k,g} - h_{g}(\alpha) \xrightarrow[N \rightarrow \infty]{a.s.} 0,
\end{align}
where $m_{g}$ is obtained as the unique solution of (\ref{fixedpoint}) and $h_{g}=\frac{1}{K} d_{g} \left( \sum_{j=1}^{S_{g}} \frac{\bar{s}_{g,j} \delta_{g}(\textbf{R}_{BS_{g}}\textbf{Q}_{g} \textbf{R}_{BS_{g}},\alpha)}{1+\frac{K_{g}}{K} d_{g}^{2} \bar{s}_{g,j} \bar{m}_{g}(\alpha) \delta_{g}(\bar{\textbf{D}}_{g},\alpha)}\right)$.

\textit{Proof:} The proof of \textit{Lemma 2} is provided in Appendix C. 

Theorem 1 and Lemma 2 approximate quantities with one occurrence of the resolvent matrix. However, with RZF, random terms involving two resolvents arise in the expression of the SINR, a case which is out of the scope of Theorem 1. We therefore develop the result for this case.

\textit{Theorem 2:} Define $\boldsymbol{\chi}(\alpha)=[\chi_{1}(\alpha),  \dots, \chi_{G}(\alpha)]^{T}$, where $\chi_{g}(\alpha)$ =$\frac{1}{K^{2}} \text{tr } \boldsymbol{\Phi}_{g} \hat{\textbf{C}}^{-1}(\alpha) \hat{\textbf{H}}^{H} \textbf{P} \hat{\textbf{H}} \hat{\textbf{C}}^{-1}(\alpha)$, where $ \boldsymbol{\Phi}_{g}$ is given by (\ref{MMSE_estimatecorr}).  Under the setting of assumptions \textbf{A-1}, \textbf{A-2} and \textbf{A-3} and for $\alpha>0$,
\begin{align}
\label{chi}
&\boldsymbol{\chi}(\alpha)- (\textbf{I}_{N}-\bar{\textbf{J}}(\alpha))^{-1} \bar{\textbf{v}}(\alpha) \xrightarrow[N \rightarrow \infty]{a.s.} 0, 
\end{align}
where,
\begin{align}
\label{J_eq}
&[\bar{\textbf{J}}(\alpha)]_{g,i}=d_{g}^{2} \frac{K_{i}}{K} \frac{1}{(1+m_{i}(\alpha))^2} \left( \bar{\beta}_{g,i} (\textbf{R}_{BS_{g}}\textbf{Q}_{g}, \textbf{Q}_{g}^{H}\textbf{R}_{BS}^{H},\alpha) + \frac{1}{\rho_{tr}} \tilde{\beta}_{i}(\textbf{R}_{BS_{g}}\textbf{Q}_{g},\textbf{Q}_{g}^{H}\textbf{R}_{BS}^{H},\alpha) \right) \\
&\bar{\textbf{v}}_{g}(\alpha) =d_{g}^{2} \frac{1}{K} \sum_{i=1}^{G} \sum_{l=1}^{K_{i}} \frac{p_{l,i}}{(1+m_{i}(\alpha))^2} \left( \bar{\beta}_{g,i} (\textbf{R}_{BS_{g}}\textbf{Q}_{g},\textbf{Q}_{g}^{H}\textbf{R}_{BS}^{H},\alpha) + \frac{1}{\rho_{tr}} \tilde{\beta}_{i}(\textbf{R}_{BS_{g}}\textbf{Q}_{g},\textbf{Q}_{g}^{H}\textbf{R}_{BS}^{H},\alpha) \right), \nonumber \\
&\text{where } \hspace{.03in} \bar{\beta}_{g,i}(\textbf{A},\textbf{B},\alpha)=  \bar{\beta}^{1}_{g,i}(\textbf{A},\textbf{B},\alpha)+\bar{\beta}^{2}_{g,i}(\textbf{A},\textbf{B},\alpha), \\
&\bar{\beta}^{1}_{g,i}(\textbf{A},\textbf{B},\alpha)=\begin{cases}  \frac{ d_{i}^{2} }{S_{i} K} \sum_{j=1}^{S_{i}} \frac{ \bar{s}_{i,j}}{(1+\frac{K_{i}}{K} d_{i}^2 \bar{m}_{i}(\alpha) \bar{s}_{i,j} \delta_{i}(\bar{\textbf{D}}_{i},\alpha))^2} \sum_{n=1}^{S_{g}}  \frac{\bar{s}_{g,n} u'_{g}(\textbf{A}\textbf{R}_{BS_{g}} \textbf{B}, \bar{\textbf{D}}_{i},\alpha)}{ \left(1+\frac{K_{g}}{K} d_{g}^2 \bar{m}_{g}(\alpha) \bar{s}_{g,n} \delta_{g}(\bar{\textbf{D}}_{g},\alpha)\right)^2}, & \text{ if } g \neq i, \\
 \frac{ d_{i}^{2} }{K} \sum_{j=1}^{S_{i}}  \frac{ \bar{s}_{i,j}}{\left(1+\frac{K_{i}}{K} d_{i}^2 \bar{m}_{i}(\alpha) \bar{s}_{i,j} \delta_{i}(\bar{\textbf{D}}_{i},\alpha)\right)^2} \Big( \underset{n \neq j}{\sum_{n=1}^{S_{i}}}  \frac{\frac{1}{S_{i}} \bar{s}_{i,n} u'_{i}(\textbf{A}\textbf{R}_{BS_{i}} \textbf{B}, \bar{\textbf{D}}_{i},\alpha)}{ \left(1+\frac{K_{i}}{K} d_{i}^2 \bar{m}_{i}(\alpha) \bar{s}_{i,n} \delta_{i}(\bar{\textbf{D}}_{i},\alpha)\right)^2} \nonumber \\
+ \bar{s}_{i,j} \delta_{i}(\textbf{R}_{BS_{i}}\textbf{Q}_{i} \textbf{R}_{BS_{i}} \textbf{B},\alpha) \delta_{i}(\textbf{A} \textbf{R}_{BS_{i}}\textbf{Q}^{H}_{i} \textbf{R}^{H}_{BS_{i}},\alpha ) \Big),& \text{ if } g = i, \nonumber
\end{cases} 
\end{align}
\begin{align}
&\bar{\beta}^{2}_{g,i}(\textbf{A},\textbf{B},\alpha)= \frac{d_{i}^{2}}{K \rho_{tr}} \sum_{j=1}^{S_{g}} \frac{ \bar{s}_{g,j} u'_{g}(\textbf{A}\textbf{R}_{BS_{g}} \textbf{B}, \tilde{\textbf{D}}_{i},\alpha)}{(1+\frac{K_{g}}{K} d_{g}^2 \bar{m}_{g}(\alpha) \bar{s}_{g,j} \delta_{g}(\bar{\textbf{D}}_{g},\alpha))^2},\\ 
&\tilde{\beta}_{i}(\textbf{A},\textbf{B},\alpha)=\frac{d_{i}^2}{K} \left(\sum_{j=1}^{S_{i}} \frac{\bar{s}_{i,j} u'_{i}(\bar{\textbf{D}}_{i}, \textbf{A}\textbf{B},\alpha)}{(1+\frac{K_{i}}{K} d_{i}^2 \bar{m}_{i}(\alpha) \bar{s}_{i,j} \delta_{i}(\bar{\textbf{D}}_{i},\alpha))^2} + \frac{S_{i}}{\rho_{tr}} u'_{i}(\tilde{\textbf{D}}_{i}, \textbf{A}\textbf{B},\alpha) \right),
\end{align}
where for arbitrary matrices $\textbf{R}_{g}$ and \textbf{L} having uniformly bounded spectral norm,
\begin{align}
&u'_{g}(\textbf{R}_{g}, \textbf{L},\alpha)=\frac{1}{S_{g}} \text{tr } \textbf{R}_{g} \left[\bar{\textbf{T}} \left(\sum_{z=1}^{G} \frac{\bar{\textbf{D}}_{z} ( d_{z}^2  K_{z}\bar{m}_{z}(\alpha))^2 }{S_{z}K^2}  \text{tr} (\bar{\textbf{S}}_{z} \textbf{W}_{z}(\alpha)^2 \bar{\textbf{S}}_{z}) u'_{z}(\bar{\textbf{D}}_{z}, \textbf{L},\alpha)  + \textbf{L} \right) \bar{\textbf{T}} \right], \nonumber \\
&\textbf{W}_{i}(\alpha)=\left( \textbf{I}_{S_{i}} + \frac{K_{i}}{K} d_{i}^2 \bar{m}_{i}(\alpha) \delta_{i}(\bar{\textbf{D}}_{i},\alpha) \bar{\textbf{S}}_{i} \right)^{-1},
\end{align}
and $\textbf{u}'(\bar{\textbf{D}}, \textbf{L},\alpha)=[u'_{1}(\bar{\textbf{D}}_{1}, \textbf{L},\alpha), u'_{2}(\bar{\textbf{D}}_{2}, \textbf{L},\alpha),\dots, u'_{G}(\bar{\textbf{D}}_{G}, \textbf{L},\alpha) ]^{T}$, which can be expressed as a system of linear equations as follows,
\begin{align}
&\textbf{u}'(\bar{\textbf{D}}, \textbf{L},\alpha)=(\textbf{I}_{N}-\textbf{J}(\bar{\textbf{D}},\alpha))^{-1} \textbf{v}(\bar{\textbf{D}}, \textbf{L},\alpha), \\
&[\textbf{J}(\bar{\textbf{D}},\alpha)]_{g,i}=\frac{1}{S_{g}} \text{tr }(\bar{\textbf{D}}_{g} \bar{\textbf{T}}(\alpha) \bar{\textbf{D}}_{i} \bar{\textbf{T}}(\alpha)) \left(\frac{(\bar{m}_{i}(\alpha))^2}{S_{i}} \left(\frac{K_{i}}{K}\right)^{2} d_{i}^{4} \text{tr }(\bar{\textbf{S}}_{i} \textbf{W}_{i}(\alpha)^{2} \bar{\textbf{S}}_{i})  \right), \\
&[\textbf{v}(\bar{\textbf{D}}, \textbf{L},\alpha)]_{g}=\frac{1}{S_{g}} \text{tr }(\bar{\textbf{D}}_{g}\bar{\textbf{T}}(\alpha) \textbf{L} \bar{\textbf{T}}(\alpha)),
\end{align}
for $g,i=1, \dots, G$.

\textit{Proof:} The proof of \textit{Theorem 2} is provided in Appendix D. 
\vspace{-.15in}
\subsection{Deterministic Approximation of the SINR} 
Based on the results in \textit{Theorem 1}, \textit{Lemma 2} and \textit{Theorem 2}, the deterministic equivalent of the SINR in (\ref{SINR_RZF}) can be derived and is presented in the next theorem. 

\textit{Theorem 3:} Under the setting of assumptions \textbf{A-1}, \textbf{A-2}  and \textbf{A-3} and for $\alpha>0$, the downlink SINR of user $k$ in group $g$ defined in (\ref{SINR_RZF}) converges almost surely as,
\begin{align}
\label{converge}
& \gamma_{k,g\text{RZF}} -\gamma^{o}_{k,g\text{RZF}} \xrightarrow[N \rightarrow \infty]{a.s.} 0,
\end{align}
where,
\begin{align}
\label{approx}
&\gamma^{o}_{k,g\text{RZF}} = \frac{p_{k,g} (h_{g}(\alpha))^{2}}{\Upsilon_{k,g}^{o} (1+m_{g}(\alpha))^2 +\frac{\xi^{o}(\textbf{I}_{N}, \textbf{I}_{N},\alpha)(1+m_{g}(\alpha))^{2}}{\rho}},
\end{align}
where,
\begin{align}
\label{interference}
&\Upsilon_{k,g}^{o} = \kappa^{o}_{g}(\textbf{I}_{N}, \textbf{I}_{N},\alpha)- 2\frac{h_{g}(\alpha)}{(1+m_{g}(\alpha))} d_{g}  \kappa^{o}_{g}(\textbf{R}_{BS_{g}}\textbf{Q}_{g}, \textbf{I}_{N},\alpha) + \left(\frac{h_{g}(\alpha)}{1+m_{g}(\alpha)}\right)^2 \chi_{g}^{o}(\alpha), \\
&\kappa^{o}_{g}(\textbf{A}, \textbf{B},\alpha)=\frac{1}{K} \sum_{i=1}^{G}\sum_{l=1}^{K_{i}} \frac{1}{(1+m_{i}(\alpha))^2} \bar{\beta}_{g,i} (\textbf{A}, \textbf{B},\alpha) (p_{l,i}+\chi^{o}_{i}(\alpha)), \\
&\xi^{o}(\textbf{A}, \textbf{B},\alpha)=\frac{1}{K} \sum_{i=1}^{G}\sum_{l=1}^{K_{i}} \frac{1}{(1+m_{i}(\alpha))^2} \tilde{\beta}_{i} (\textbf{A}, \textbf{B},\alpha) (p_{l,i}+\chi^{o}_{i}(\alpha)). 
\end{align}

\textit{Proof:} The proof of \textit{Theorem 3} is given in Appendix F.

\textit{Corollary 1:}  Assume that \textbf{A-1}, \textbf{A-2} and \textbf{A-3} hold true and $\alpha>0$. Then the individual downlink rates $R_{k,g}$ of users converge as, 
\begin{align}
&R_{k,g}-R_{k,g}^{o} \xrightarrow[N \rightarrow \infty]{a.s.} 0,
\end{align}
where,
\begin{align}
\label{rate}
R_{k,g}^{o}=\log(1+\gamma^{o}_{k,g\text{RZF}} ),
\end{align}
where $\gamma^{o}_{k,g\text{RZF}}$ is given by (\ref{approx}).

\textit{Proof:} The proof follows from the application of the continuous mapping theorem \cite{mapping} to the logarithm function and the almost sure convergence of $\gamma_{k,g\text{RZF}}$ in (\ref{converge}).

An approximation of the average system sum rate can be obtained by replacing $R_{k,g}$ in (\ref{summm}) with its asymptotic approximation as,
\begin{align}
\label{sum}
R^{o}_{sum}=\sum_{g=1}^{G} \sum_{k=1}^{K_{g}} \log (1+\gamma^{o}_{k,g\text{RZF}}).
\end{align}

These asymptotic expressions will be shown to provide good approximations even for moderate system sizes by the means of simulations in Section IV. This means they can be used for evaluating the performance of practical systems without relying on time-consuming Monte-Carlo simulations. These expressions  are also useful in performing different optimization tasks, for example, determining the optimal $\alpha$ and power allocation as shown in \cite{SINRdeterministic}.

The deterministic equivalents depend only on the `slowly varying' covariance matrices $\textbf{R}_{BS_{g}}$ and $\bar{\textbf{S}}_{g}$, instead of the instantaneous channels that vary very fast. These correlation matrices, although huge in size in the large $N,S$  regime, can be computed at the BS using knowledge of only the large-scale channel statistics. In particular, each correlation matrix, can be computed using estimates of the angular spreads, $\sigma_{t,g}, \sigma_{s,g}$,  and the mean AoD, $\mu_{t,g}= \mu_{s,g}$. These parameters are illustrated in Fig. \ref{scattering_model}. The spreads can be locally estimated while $\mu_{t,g}$ depends on the line of sight  (LoS) angles of the users within  a group, that have to be fed to the BS. However, the estimation of these large-scale parameters must be performed only once per coherence period (rather than at the same pace as the small-scale fading) and the number of required estimated parameters depends on $G$ and $K$ (much less than $N$ and $S$ in practical systems). Discussion on the estimation of large-scale parameters  can be found in \cite{ourworkTCOM,usergroups}, where the authors reason that the estimation of channel covariance matrices imposes a low CSI feedback overhead.

We would also point out here that the considered RZF precoding, known for its high performance, also imposes a high computational complexity in large-scale MIMO systems due to the matrix inversion operation. It is therefore interesting to study the double scattering model with other precoding schemes that reduce this prohibitively high computational complexity. One way is to employ truncated polynomial expansion (TPE) precoding proposed in  \cite{linearprecoding}, which replaces the matrix inversion by a TPE. The TPE precoding can perform close to RZF, with a reduced computational burden. Theorem 1 and Theorem 2 developed in this work will play a direct role in the asymptotic analysis of the SINR of the double scattering model with TPE. 


Despite being useful, the expressions in this work are quite involved and do not provide direct insights into the interplay between different system parameters. We therefore focus on Rayleigh product channel and obtain some simplifications under different operating conditions. 
\vspace{-.2in}
\subsection{Rayleigh Product Channel}
A special case of the double-scattering channel is the Rayleigh product channel which does not exhibit any form of correlation \cite{doublescattering4}. For this model, popularly known as the multi-keyhole channel, we find that Theorem 3 can be given in a closed-form as shown in the next corollary.

\textit{Corollary 2:}  For $G=1$, let $S_{1}=S$, $K_{1}=K$, and assume $\bar{\textbf{S}}_{1}=\textbf{I}_{S}$ and $\textbf{R}_{BS_{1}}=\textbf{I}_{N}$. Then $\gamma^{o}_{k\text{RZF}}$ in Theorem 3 can be given in a closed-form as,
\begin{align}
&\gamma^{o}_{k\text{RZF}}=\frac{\frac{p_{k}}{\bar{P}/K}a^2h^2(\rho_{tr}-\tilde{\beta} \bar{m}^2 -\bar{\beta}\bar{m}^2 \rho_{tr})}{\bar{\beta}\rho_{tr} +a^2\tilde{\beta} h^2 \bar{m}^2-2ah\bar{\beta}\bar{m}\rho_{tr}+a^2 \bar{\beta} h^2 \bar{m}^2 \rho_{tr}+\frac{\tilde{\beta} \rho_{tr}}{\rho}}, \\
&\text{where $a=\frac{1}{1+\frac{1}{\rho_{tr}}}$,  }  h=\frac{Sa(N-K+K\bar{m})}{K(S\alpha+Ka^2\bar{m}^2-Ka^2\bar{m}+Na^2\bar{m})}, \\
&\bar{\beta}= \frac{Sa^4(N-K+K\bar{m})^2}{K(S\alpha+Ka^2\bar{m}^2-Ka^2\bar{m}+Na^2\bar{m})^2} + \frac{S^4 \alpha^4 (S-1) u' }{K(S\alpha+Ka^2\bar{m}^2-Ka^2\bar{m}+Na^2\bar{m})^4} \nonumber \\
&+\frac{S^3 \alpha^2 u'}{K\rho_{tr}(S\alpha+Ka^2\bar{m}^2-Ka^2\bar{m}+Na^2\bar{m})^2}, \\
&\tilde{\beta}=\frac{Su'}{K\rho_{tr}}+\frac{S^3 \alpha^2 u'}{K(S\alpha+Ka^2\bar{m}^2-Ka^2\bar{m}+Na^2\bar{m})^2}, \\
&u'=\frac{a^4 \rho_{tr} (N-K+K\bar{m})(S\alpha+Ka^2\bar{m}^2-Ka^2\bar{m}+Na^2\bar{m})}{S\alpha(2Ka^4\bar{m}^2(\bar{m}-1)+2Na^4\bar{m}^2+S\alpha^2\rho_{tr}+Sa^2\alpha\bar{m}+a^2\alpha\bar{m}\rho_{tr}(2K\bar{m}-2K+N+S))}, \nonumber
\end{align}
and $\bar{m} \in \left(0,1 \right)$ is given as the unique root to,
\begin{align}
\label{polynomial}
& \bar{m}^4+\left(\frac{2N}{K}+\frac{\rho_{tr}\alpha}{a^2}-2\right)\bar{m}^3+ \left(\frac{N^2}{K^2}+1+\frac{S\alpha(1+\rho_{tr})}{Ka^2}-\frac{2N}{K}+\frac{\alpha\rho_{tr}}{a^2}\left(\frac{N}{K}-2\right) \right)\bar{m}^2 \nonumber \\
&+ \left(\frac{S\alpha^2\rho_{tr}}{Ka^4}+\frac{NS\alpha(1+\rho_{tr})}{K^2a^2} - \frac{S\alpha(1+\rho_{tr})}{Ka^2} +\frac{\alpha\rho_{tr}}{a^2}\left(1-\frac{N}{K}\right)\right) \bar{m} -\frac{\rho_{tr}S\alpha^2}{Ka^4}=0.
\end{align}

\textit{Sketch of Proof:} One can show by straightforward but tedious calculations that the fundamental equations in (\ref{fixedpoint}) can be reduced to a single polynomial equation (\ref{polynomial}) in $\bar{m}$, for a single group and $\bar{\textbf{S}}_{1}=\textbf{I}_{S_{1}}$ and $\textbf{R}_{BS_{1}}=\textbf{I}_{N}$. Simplifying (\ref{F_exp}) for the setting of Corollary 2, writing it as a polynomial in $F(\textbf{R},\textbf{L})$ and applying implicit function theorem with respect to $l$ yields the expression of $u'$. After some tedious calculations, we show that Theorem 2 can be given in a closed form and the SINR expressions in Theorem 3 can be greatly simplified.

\textit{Corollary 3:}  Under the setting of Corollary 2, let $\frac{S}{N}, \frac{S}{K} \rightarrow \infty$. Then $\gamma^{o}_{k\text{RZF}}$ defined in Theorem 3 approaches the limit $\gamma^{o}_{k\text{RZF},\frac{S}{N}, \frac{S}{K} \rightarrow\infty}$ which is given by,
\begin{align}
\label{asym_SINR}
& \gamma^{o}_{k\text{RZF},\frac{S}{N}, \frac{S}{K} \rightarrow\infty}=\frac{\frac{p_{k}}{\bar{P}/K}(a^2h^2(\rho_{tr}-\bar{\beta}\bar{m}^2(1+\rho_{tr}))}{\bar{\beta}\rho_{tr}\left(1+\frac{1}{\rho}\right)+a^2h^2\bar{m}^2\bar{\beta}(1+\rho_{tr})-2ah\bar{m}\rho_{tr}\bar{\beta}}, 
\end{align}
where $h=\frac{a}{\alpha}\left(\frac{N}{K}-1+\bar{m}\right)$, $\bar{\beta}=\frac{a^4\left(\frac{N}{K}-1+\bar{m}\right)(1+\rho_{tr})}{\alpha(\alpha\rho_{tr}+a^2\bar{m}+a^2\bar{m}\rho_{tr})}$, $\bar{m}=\frac{1-\frac{N}{K}-c+\sqrt{(c+\frac{N}{K}-1)^2+4c}}{2}$ and $c=\frac{\rho_{tr}\alpha}{a^2(1+\rho_{tr})}$.

Note that as $\frac{S}{N}, \frac{S}{K} \rightarrow \infty$, \textbf{h} behaves as a Rayleigh fading channel, whose SINR is given as (\ref{asym_SINR}). Setting, $\rho_{tr}\rightarrow \infty$, we obtain the following corollary for the perfect CSI case.

\textit{Corollary 4:}  Under the setting of Corollary 3, let $\rho_{tr} \rightarrow \infty$. Then $\gamma^{o}_{k\text{RZF},\frac{S}{N}, \frac{S}{K} \rightarrow\infty}$ is given by,
\begin{align}
\label{asym_SINR1}
& \gamma^{o}_{k\text{RZF},\frac{S}{N}, \frac{S}{K} \rightarrow\infty}=\frac{\frac{p_{k}}{\bar{P}/K}(\frac{1}{\bar{m}}-1)(1+\frac{\alpha}{\bar{m}^{2}})}{1+\frac{1}{\rho \bar{m}^{2}}},
\end{align}
where $\bar{m}=\frac{1-\frac{N}{K}-\alpha+\sqrt{(\alpha+\frac{N}{K}-1)^2+4\alpha}}{2}$.

This result has also been obtained in Corollary 2 of \cite{SINRdeterministic}, where the authors derive the deterministic equivalent of the SINR under RZF, with the channels modeled as correlated Rayleigh.

\textit{Corollary 5:}  Under the setting of Corollary 2, let $\frac{N}{S}\rightarrow \infty$ and $\frac{N}{K}\rightarrow \infty$ with $K>S$. Then $\gamma^{o}_{k\text{RZF}}$ defined in Theorem 3  approaches the limit $\gamma^{o}_{k\text{RZF},\frac{N}{S}, \frac{N}{K}\rightarrow \infty}$ which is given by,
\begin{align}
\label{asym_SINR2}
& \gamma^{o}_{k\text{RZF}, \frac{N}{S}, \frac{N}{K}\rightarrow \infty}=\frac{\frac{p_{k}}{\bar{P}/K}S}{(K-S)}.
\end{align}

\textit{Corollary 6:} If $K\leq S$ in the setting of Corollary 5, then $\bar{m} \rightarrow 0$ and as a consequence, the interference term $\Upsilon_{k}^{o} \rightarrow 0$ and the power normalization term $\xi^{o}(\textbf{I}_{N},\textbf{I}_{N})  \rightarrow 0 $ in Theorem 3. 

\textit{Remark 1:} Corollary 6 implies for $\frac{N}{S},\frac{N}{K} \rightarrow \infty$ with $K\leq S$, $\gamma^{o}_{k\text{RZF}, \frac{N}{S}, \frac{N}{K}\rightarrow \infty}$ grows unboundedly.


Two very important observations are in order now. First, the performance of a large-scale MIMO system is limited by the number of scatterers in the propagation environment. Recent works that show the spectral efficiency of large-scale MIMO systems to be unlimited base their analysis on correlated Rayleigh channels and neglect the impact of limited scattering, that may exist in practice. Even though some of these works realize that poor scattering will deteriorate the performance but they do not account for it in the theoretical analysis. This is the first paper that derives a bound on the asymptotic ergodic rate that can be achieved under limited scattering. 

The second observation is that the maximum number of users that can be served simultaneously while achieving large-scale MIMO gains is less than or equal to the number of scatterers. In fact, we find through Corollary 6 that if $K\leq S$, $ \gamma^{o}_{k\text{RZF}, \frac{N}{S}, \frac{N}{K}\rightarrow \infty}$ grows unboundedly with $N$. This `large-scale MIMO gain' is in accordance with the results in \cite{massiveMIMOO} and \cite{massive2}, which when simplified for a single cell case, reveal that $ \gamma_{k\text{RZF}} \rightarrow \infty$ for $N \rightarrow \infty$, $K/N \rightarrow 0$. The inherent assumption in these works that rely on the correlated Rayleigh  channel is that there is enough scattering to serve all the $K$ users. However, as $K$ exceeds $S$, we start to see a significant performance loss.  For example, with uniform power allocation, serving $K=S+1$ users would result in the SINR of each user approaching $S$ as $N\rightarrow \infty$, according to  Corollary 5. If we serve $K=S+2$ users, the SINR will approach $S/2$, resulting in a $3\rm{dB}$  loss. Thus, the value of $S$ is what determines the maximum number of users that can be served while realizing large-scale MIMO gains.  

\vspace{-.15in}
\section{Simulation Results}
\vspace{-.05in}

Under the double scattering model, the correlation matrices $\textbf{R}_{BS_{g}}$ and $\bar{\textbf{S}}_{g}$ are given as $\textbf{R}_{BS_{g}}=\textbf{G}(\mu_{t,g}, \sigma_{t,g}, d_{t}, S_{g})$ and $\bar{\textbf{S}}_{g}=\textbf{G}(\mu_{s,g}, \sigma_{s,g}, d_{s,g}, S_{g})$, where $\textbf{G}(\mu, \sigma, d, n)$ is defined as \cite{doublescattering},
\begin{align}
&[\textbf{G}(\mu, \sigma, d, n)]_{k,l}=\frac{1}{n} \sum_{j=\frac{1-n}{2}}^{\frac{n-1}{2}} \exp \Big(-i 2\pi d (k-l) \cos \Big(\frac{\pi}{2}+\frac{j\sigma}{n-1} + \mu \Big)  \Big).
\end{align}

\begin{figure}[!t]
\begin{minipage}[b]{0.45\linewidth}
\centering
\includegraphics[width=2.65 in]{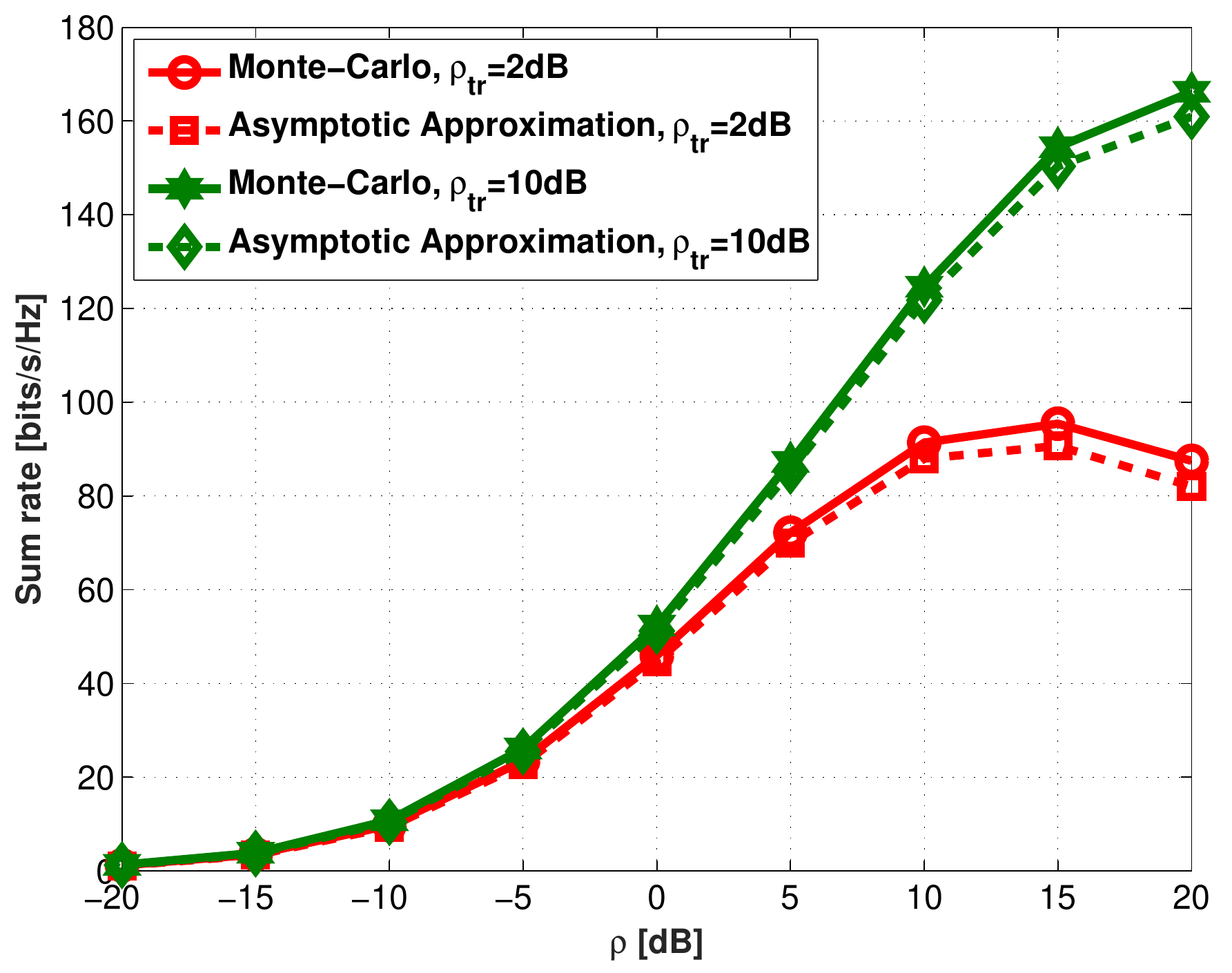}
\caption{Sum rate versus SNR with $\alpha=1/\rho$.}
\label{validation}
\end{minipage}
\hspace{0.5cm}
\begin{minipage}[b]{0.45\linewidth}
\centering
\includegraphics[width=2.65 in]{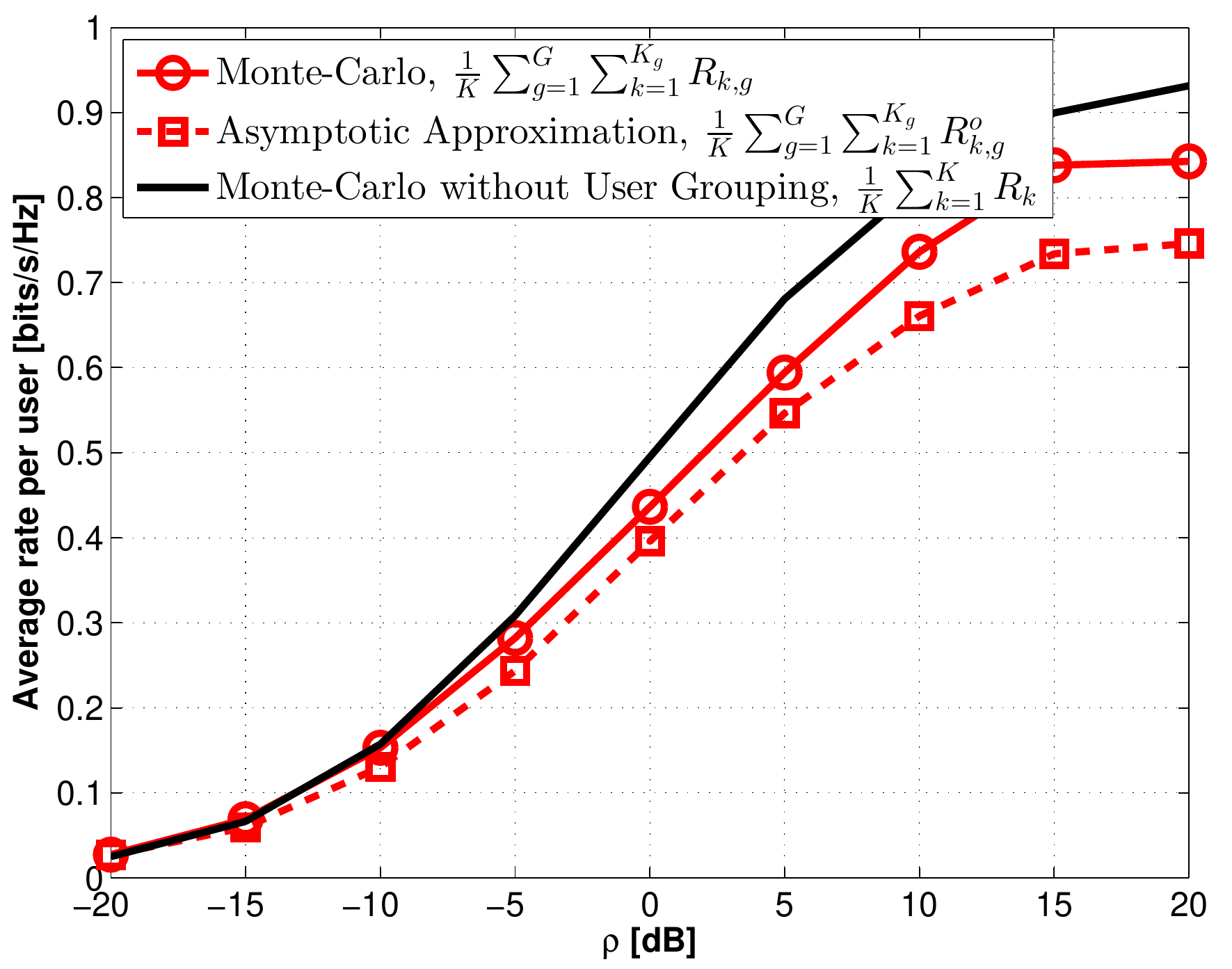}
\caption{Average rate for $\rho_{tr}=2\rm{dB}$, $\alpha=1/\rho$. }
\label{wogroup}
\end{minipage}
\end{figure}

All parameters have already been defined in Fig. \ref{scattering_model}. The parameter values are set as $G=4$, $K=128$, $N=128$, $S_{g}$=$\{130, 140, 134, 144 \}$, $\mu_{t,g}=\mu_{s,g}$=$\{-\pi/3,-\pi/9, \pi/9, \pi/3  \}$,  $\sigma_{t,g}$=$\{ \pi/5, \pi/6, \\ \pi/5, \pi/7 \}$ and $\sigma_{s,g}$=$\{ \pi/6, \pi/6, \pi/6, \pi/6 \}$. Also, $d_{t}=0.5$ and $d_{s,g}=2$ for all $g$. We assume an equal number of users in each group, i.e. $K_{g}=K/G$, with uniform power allocation, $\textbf{P}=\textbf{I}_{K}$. Fig. \ref{validation} compares the downlink system sum rate $R_{sum}=\sum_{g=1}^{G}\sum_{k=1}^{K_{g}}  \log(1+\gamma_{{k,g}_{\text{RZF}}})$ obtained using 2000 Monte-Carlo realizations of the SINR in (\ref{SINR_RZF}) to the asymptotic approximation provided in (\ref{sum}), where $\gamma^{o}_{k,g\text{RZF}}$ is given by (\ref{approx}).  It can be seen that the asymptotic result yields a very good approximation for moderate system dimensions.

 To explain the decrease in sum-rate at high values of $\rho$ for $\rho_{tr}=2\rm{dB}$, note that the goal of the regularization parameter $\alpha$ in the RZF precoder in (\ref{RZF}) is to improve the conditioning of the random matrix $\hat{\textbf{H}}^{H}\hat{\textbf{H}}+K\alpha \textbf{I}_{N}$ and hence it should depend on the channel parameters, including the training SNR $\rho_{tr}$ and the downlink SNR $\rho$. Finding the optimal $\alpha$ that maximizes $\gamma^{o}_{k,g\text{RZF}}$ is outside the scope of this work. Now for the chosen $\alpha=1/\rho$ (optimal under perfect CSI), the sum rate is decreasing at high SNR values for $\rho_{tr}=2\rm{dB}$. This is because $\alpha$ does not account for $\rho_{tr}$ and thus the matrix $\hat{\textbf{H}}^{H}\hat{\textbf{H}}+K\alpha \textbf{I}_{N}$ in the RZF precoder becomes ill-conditioned as the quality of the estimate deteriorates and $\alpha$ decreases (due to increase in $\rho$). This results in a loss in the performance. This effect has also been observed in  (\cite{SINRdeterministic}, Fig. 2). In fact, the authors there obtain the optimal $\alpha^{*}$ as a solution of a fixed point equation that depends on $\rho_{tr}$ and show that for $\alpha=\alpha^{*}$, the sum rate continues to increase with $\rho$ for any quality of channel estimate.

 Furthermore, note that the mismatch between the theoretical and the Monte-Carlo results starts to increase for high SNR values due to the slower convergence of $\gamma_{k,g\text{RZF}}$ to its deterministic approximation as well documented in RMT literature \cite{SINRdeterministic,convergence}. A better approximation at high SNR can be seen in systems with higher values of $N,K$ and $S$. 

Next we plot in Fig. \ref{wogroup} the average achievable rate for $G=11$, $N=121$, $S_{g}=120$ $\forall g$, and $K=60$ users having mean AoDs generated between $-15^{o}$ to $15^{o}$ and spreads between $3^o$ and $6^o$. We employ the fixed quantization method from \cite{usergroups} to group the users, based on the criteria of the minimum chordal distance between the users' correlation eigenspaces and the group subspaces. The Monte-Carlo and deterministic average rates when common correlation matrices are assumed are shown to be close. In fact, the average (and sum) rate now continues to increase for $\rho_{tr}=2\rm{dB}$ because for $K<N$, $\hat{\textbf{H}}^{H}\hat{\textbf{H}}+K\alpha \textbf{I}_{N}$ is  well-conditioned.  The Monte-Carlo average rate under the actual different correlation matrices of the users is plotted in black to highlight the importance of having variation in correlation matrices. However, for us to analyze this difficult non-Gaussian model, it was necessary to assume common correlation matrices within groups. The performance loss due to this assumption can be reduced by increasing the number of groups. 



\begin{figure}[!t]
\begin{minipage}[b]{0.45\linewidth}
\centering
\includegraphics[width=2.65 in]{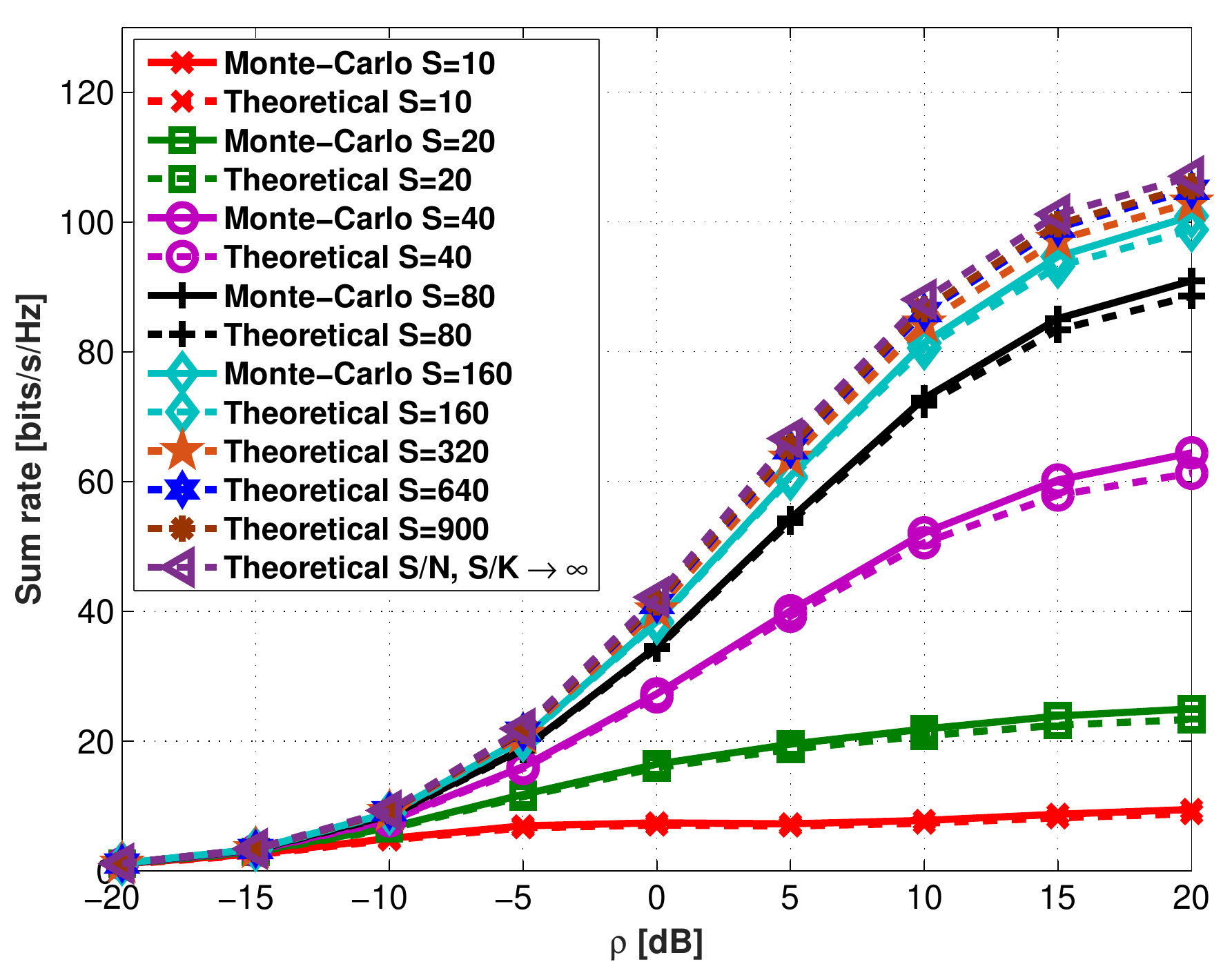}
\caption{Sum rate for a multi-keyhole channel.}
\label{withS}
\end{minipage}
\hspace{0.5cm}
\begin{minipage}[b]{0.45\linewidth}
\centering
\includegraphics[width=2.65 in]{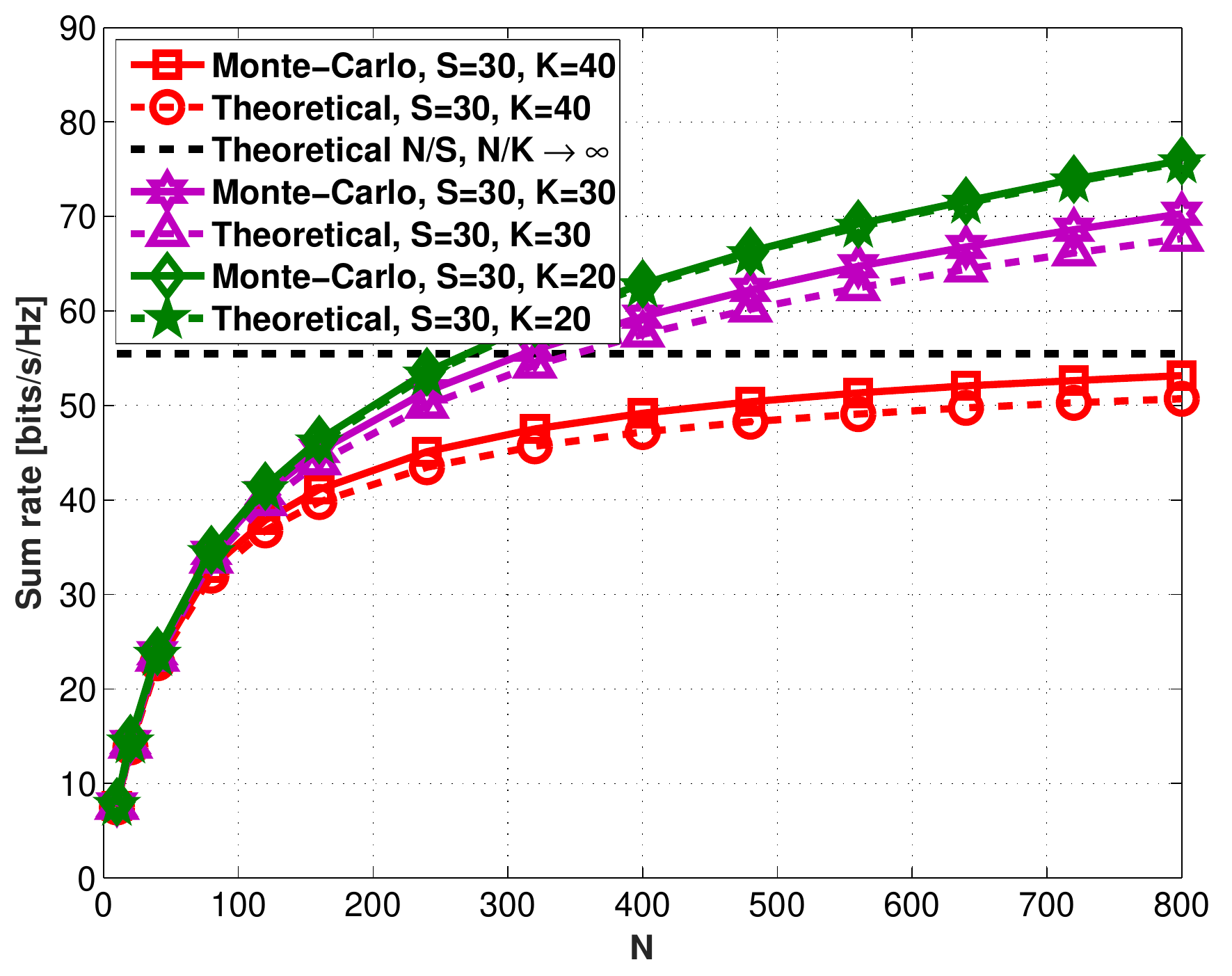}
\caption{Sum rate versus $N$ with $\rho=5\rm{dB}$, $\alpha=1/\rho$.}
\label{withNBS}
\end{minipage}
\end{figure}

Fig. \ref{withS} studies the effect of the number of scatterers on the system sum rate for a single group with multi-keyhole channel, i.e. $G=1$, $\textbf{R}_{BS}=\textbf{I}_{N}$, $\bar{\textbf{S}}=\textbf{I}_{S}$, with $N=128$, $K=32$, $\alpha=1/\rho$ and $\rho_{tr}=10\rm{dB}$. The downlink sum rate in (\ref{sum}) plotted using the closed-form expression of $\gamma^{o}_{k\text{RZF}}$ in Corollary 2 is close to the Monte-Carlo result even  for a very low number of scatterers. The spatial multiplexing gains are seen to increase linearly with $S$. However, for $S>N$, the gains start to decrease since the degrees of freedom are limited by the number of antennas at the BS. The limiting sum rate  as $S/N, S/K \rightarrow \infty$ is also plotted using the SINR in (\ref{asym_SINR}). As the number of scatterers increases, the performance approaches to that of a Rayleigh fading channel.

Finally, we study the performance of the Rayleigh product channel in Fig. \ref{withNBS} as the number of BS antennas increases for a single group.  It can be seen that the performance for the case when $S=30$, $K=40$ saturates to a limiting sum rate, given by substituting (\ref{asym_SINR2}) in (\ref{sum}),  which is also plotted in black on the figure. This result confirms that it is useless to deploy more antennas when the number of scatterers in the environment is limited. The other curves highlight that when $K\leq S$, we see the `large-scale MIMO effect' and the performance continues to grow with $N$ because the channel has enough degrees of freedom to support all the users.  
\vspace{-.15in}
\section{Conclusion}
\vspace{-.06in}
In this paper, we studied a large-scale multi-user MISO system with double-scattering channels that are more realistic than the commonly used Gaussian channels. We first derived the MMSE estimate for this channel. Then under the assumption of per-group channel correlation matrices, we derived the deterministic approximations of the SINR and ergodic rates with RZF precoding, that are tight in the large system limit. Simulation results showed a close match between the asymptotic and the Monte-Carlo simulated sum rate   and provided insights into the performance of multi-keyhole channels. We showed that large-scale MIMO gains can only be realized when the number of scheduled users is less than the number of scatterers in the environment.

\appendices
\vspace{-.1in}
\section{Proof of Lemma 1}
\vspace{-.05in}
The MMSE estimate $\hat{\textbf{h}}_{k,g}$ of $\textbf{h}_{k,g}$ is computed as $\hat{\textbf{h}}_{k,g}=\textbf{F}_{g} \textbf{y}^{tr}_{k,g}$ where $\textbf{F}_{g}$ is obtained as the solution of $\text{min }_{\textbf{F}_{g}} \mathbb{E}[||\textbf{h}_{k,g}-\textbf{F}_{g}\textbf{y}^{tr}_{k,g}||^2]$ resulting in,
\vspace{-.05in}
\begin{align}
&\textbf{F}_{g}=\textbf{C}_{\textbf{h}_{k,g}\textbf{y}^{tr}_{k,g}} \textbf{C}^{-1}_{\textbf{y}^{tr}_{k,g}\textbf{y}^{tr}_{k,g}}.
\end{align}
Using the expression of $\textbf{y}_{k,g}^{tr}$ in (\ref{y_tr}) and the independence of $\textbf{h}_{k,g}$ and $\textbf{n}_{k,g}$ we get,
\begin{align}
&\textbf{C}_{\textbf{h}_{k,g}\textbf{y}^{tr}_{k,g}}=\mathbb{E}[\textbf{h}_{k,g}\textbf{y}^{tr^{H}}_{k,g}] = \mathbb{E}[\textbf{h}_{k,g}\textbf{h}_{k,g}^{H}]=\mathbb{E}[\textbf{R}_{BS_{g}}^{1/2} \textbf{W}_{g}\bar{\textbf{S}}_{g}^{1/2} \tilde{\textbf{w}}_{k,g} \tilde{\textbf{w}}_{k,g}^{H} \bar{\textbf{S}}_{g}^{1/2} \textbf{W}_{g}^{H} \textbf{R}_{BS_{g}}^{1/2^{H}}]. 
\end{align}
Conditioning the expectation on $\textbf{W}_{g}$ first, we obtain,
\begin{align}
&\textbf{C}_{\textbf{h}_{k,g}\textbf{y}^{tr}_{k,g}}=\textbf{R}_{BS_{g}}^{1/2} \mathbb{E}[\textbf{W}_{g}\bar{\textbf{S}}_{g}^{1/2} \mathbb{E}[\tilde{\textbf{w}}_{k,g} \tilde{\textbf{w}}_{k,g}^{H}|\textbf{W}_{g}] \bar{\textbf{S}}_{g}^{1/2} \textbf{W}_{g}^{H}] \textbf{R}_{BS_{g}}^{1/2^{H}}=\frac{1}{S_{g}} \textbf{R}_{BS_{g}}^{1/2} \mathbb{E}[\textbf{W}_{g}\bar{\textbf{S}}_{g} \textbf{W}_{g}^{H}] \textbf{R}_{BS_{g}}^{1/2^{H}}, \\
&=\frac{1}{S_{g}} \textbf{R}_{BS_{g}}^{1/2} (\text{tr } \bar{\textbf{S}}_{g}) \textbf{I}_{N} \textbf{R}_{BS_{g}}^{1/2^{H}}=\frac{1}{S_{g}} (\text{tr } \bar{\textbf{S}}_{g}) \textbf{R}_{BS_{g}}.\\
&\text{Similarly, } \textbf{C}_{\textbf{y}^{tr}_{k,g}\textbf{y}^{tr}_{k,g}}=\mathbb{E}[\textbf{y}^{tr}_{k,g}\textbf{y}^{tr^{H}}_{k,g}]=\mathbb{E}\left[\textbf{h}_{k,g}\textbf{h}_{k,g}^{H}+\frac{1}{\rho_{tr}} \textbf{n}^{tr}_{k,g}\textbf{n}^{tr^{H}}_{k,g}\right]=\frac{(\text{tr } \bar{\textbf{S}}_{g})}{S_{g}} \textbf{R}_{BS_{g}} + \frac{1}{\rho_{tr}} \textbf{I}_{N}.
\end{align}
Therefore $\hat{\textbf{h}}_{k,g}= \frac{1}{S_{g}} (\text{tr } \bar{\textbf{S}}_{g})\textbf{R}_{BS_{g}} \textbf{Q}_{g}\textbf{y}^{tr}_{k,g}$, where $\textbf{Q}_{g}=\left(\frac{1}{S_{g}} (\text{tr } \bar{\textbf{S}}_{g}) \textbf{R}_{BS_{g}} + \frac{1}{\rho_{tr}} \textbf{I}_{N}\right)^{-1}$.

\vspace{-.15in}
\section{Proof of Theorem 1}
As a starting point we assume $\textbf{Z}_{g}$ to be deterministic such that $\text{lim sup}_{N}||\textbf{Z}_{g}\textbf{Z}_{g}^{H}||< \infty$. Under this setting, we use the expression of $\hat{\textbf{h}}_{k,g}$ given by (\ref{MMSE_estimate1}) in the resolvent matrix, $\hat{\textbf{C}}^{-1}$, to have,
\begin{align}
\label{C}
&\hat{\textbf{C}}=\frac{1}{K} \sum_{g=1}^{G} \sum_{k=1}^{K_{g}}  \boldsymbol{\Phi}^{1/2}_{g} \bar{\textbf{q}}_{k,g} \bar{\textbf{q}}_{k,g}^{H} \boldsymbol{\Phi}^{1/2^{H}}_{g} + \alpha \textbf{I}_{N}=\sum_{g=1}^{G} \tilde{ \boldsymbol{\Phi}}^{1/2}_{g} \bar{\textbf{Q}}_{g} \textbf{I}_{K_{g}} \bar{\textbf{Q}}_{g}^{H} \tilde{\boldsymbol{\Phi}}^{1/2^{H}}_{g}+ \alpha \textbf{I}_{N}, 
\end{align}
where the entries of $\bar{\textbf{Q}}_{g} \in \mathbb{C}^{N \times K_{g}} \sim \mathcal{CN}(0, \frac{1}{K_{g}})$ and $\tilde{ \boldsymbol{\Phi}}_{g}=\frac{K_{g}}{K} \boldsymbol{\Phi}_{g}$, where $\boldsymbol{\Phi}_{g}$ is given by (\ref{MMSE_estimatecorr}).  Note that $\tilde{ \boldsymbol{\Phi}}^{1/2}_{g} \bar{\textbf{Q}}_{g} \textbf{I}_{K_{g}} \bar{\textbf{Q}}_{g}^{H} \tilde{\boldsymbol{\Phi}}^{1/2^{H}}_{g}$ is a double scattering model with deterministic $\tilde{ \boldsymbol{\Phi}}^{1/2}_{g}$s. Under this setting, the deterministic equivalent of $\frac{1}{K} \text{tr }\textbf{U} \hat{\textbf{C}}^{-1}$ is obtained using Corollary 1 from \cite{theorem} as,
\begin{align}
\label{ref1}
&\frac{1}{K} \text{tr }\textbf{U} \hat{\textbf{C}}^{-1} - \frac{1}{K} \text{tr }\textbf{U}  \left(\sum_{i=1}^{G} \bar{e}_{i} \tilde{ \boldsymbol{\Phi}}_{i}  + \alpha \textbf{I}_{N}\right)^{-1} \xrightarrow[N \rightarrow \infty]{a.s.} 0,
\end{align}
where $(e_{g}, \bar{e}_{g})$ are given as a unique solution to the following set of implicit equations,
\begin{align}
\label{e_quant}
&\bar{e}_{g}=\frac{1}{1+e_{g}}, 
\end{align}
\begin{align}
\label{e_quant1}
&e_{g}= \frac{1}{K_{g}} \text{tr }\tilde{ \boldsymbol{\Phi}}_{g}  \left(\sum_{i=1}^{G} \bar{e}_{i}\tilde{ \boldsymbol{\Phi}}_{i}  + \alpha \textbf{I}_{N}\right)^{-1}.
\end{align}

Now for the actual double-scattering channel model, $\textbf{Z}_{g}$s are random and modeled as \\ $\frac{1}{\sqrt{S_{g}}}\textbf{R}_{BS_{g}}^{1/2}\textbf{W}_{g}\bar{\textbf{S}}_{g}^{1/2}$. Using the Fubini Theorem \cite{iterative} we can extend the result in (\ref{ref1}) to random $\textbf{Z}_{g}$s by finding deterministic equivalents, denoted as $m_{g}$ and $\bar{m}_{g}$, for $e_{g}$ and $\bar{e}_{g}$ respectively. 

To do this, we first define quantities $e_{g,i,j}$ and $\bar{e}_{g,i,j}$ for $i=1,\dots, G$, $j=1,\dots, S_{i}$ as $\bar{e}_{g,i,j}=\frac{1}{1+e_{g,i,j}}$ and $e_{g,i,j}= \frac{1}{K_{g}} \text{tr }\tilde{ \boldsymbol{\Phi}}_{g,i,j}  (\sum_{l=1}^{G} \bar{e}_{l,i,j}\tilde{ \boldsymbol{\Phi}}_{l,i,j}  + \alpha \textbf{I}_{N})^{-1}$, where,
\begin{align}
&\tilde{ \boldsymbol{\Phi}}_{g,i,j}=\frac{K_{g}}{K} d_{g}^{2} \textbf{R}_{BS_{g}} \textbf{Q}_{g} \left( \textbf{Z}_{g,i,j} \textbf{Z}^{H}_{g,i,j} + \frac{1}{\rho_{tr}} \textbf{I}_{N}\right)  \textbf{Q}^{H}_{g} \textbf{R}^{H}_{BS_{g}}, \\
&\textbf{Z}_{g,i,j}=
\begin{cases}
\textbf{Z}_{g}, & \text{if $i \neq g$}, \\
\small [\textbf{z}_{g,1}, \dots, \textbf{z}_{g,j-1}, \textbf{z}_{g,j+1}, \dots, \textbf{z}_{g,S_{g}}], & \text{if $i=g$}.
\end{cases} 
\end{align}
\normalsize
%
It can be shown following the techniques used in Appendix E of \cite{iterative} that for all $g,i,j$,
\begin{align}
\label{cond_e}
&e_{g}-e_{g,i,j} \xrightarrow[K \rightarrow \infty]{a.s.} 0, \\
\label{cond_e1}
&\bar{e}_{g}-\bar{e}_{g,i,j} \xrightarrow[K \rightarrow \infty]{a.s.} 0.
\end{align} 

Now using the expression of $\boldsymbol{\Phi}_{g}$ from (\ref{MMSE_estimatecorr}) in the expression of $e_{g}$ in (\ref{e_quant1}) we have, 
\begin{align}
&e_{g}= \frac{d_{g}^{2} }{K} \text{tr } \textbf{R}_{BS_{g}} \textbf{Q}_{g} \left( \textbf{Z}_{g} \textbf{Z}_{g}^{H} + \frac{1}{\rho_{tr}} \textbf{I}_{N} \right)\textbf{Q}^{H}_{g} \textbf{R}^{H}_{BS_{g}} \Big(\sum_{i=1}^{G}  \bar{e}_{i} \tilde{\boldsymbol{\Phi}}_{i} + \alpha \textbf{I}_{N} \Big)^{-1}, 
\end{align}
\begin{align}
&= \frac{d_{g}^2}{K}  \sum_{j=1}^{S_{g}} \textbf{z}_{g,j}^{H} \textbf{Q}^{H}_{g} \textbf{R}^{H}_{BS_{g}} \left(\sum_{i=1}^{G}  \bar{e}_{i} \tilde{\boldsymbol{\Phi}}_{i} + \alpha \textbf{I}_{N} \right)^{-1} \textbf{R}_{BS_{g}} \textbf{Q}_{g} \textbf{z}_{g,j} +\frac{d_{g}^2 }{K} \frac{1}{\rho_{tr}} \text{tr } \tilde{\textbf{D}}_{g} \left(\sum_{i=1}^{G}  \bar{e}_{i} \tilde{\boldsymbol{\Phi}}_{i} + \alpha \textbf{I}_{N} \right)^{-1}, \nonumber \\
\label{T_ex}
&= T_{1}+T_{2},
\end{align}
where $\tilde{\textbf{D}}_{g}=\textbf{R}_{BS_{g}} \textbf{Q}_{g} \textbf{Q}^{H}_{g} \textbf{R}^{H}_{BS_{g}}$. First we treat $T_{1}$ using (\ref{cond_e1}) and matrix inversion lemma (Lemma 1 in \cite{massiveMIMOO}) to remove the dependence of $\left(\sum_{i=1}^{G}  \bar{e}_{i} \tilde{\boldsymbol{\Phi}}_{i} + \alpha \textbf{I}_{N} \right)^{-1}$ on the vector $\textbf{z}_{g,j}$ as,
\begin{align}
&T_{1}=\frac{d_{g}^2}{K}  \sum_{j=1}^{S_{g}} \left(\frac{\splitfrac{\textbf{z}_{g,j}^{H} \textbf{Q}^{H}_{g} \textbf{R}^{H}_{BS_{g}} \Big(\sum_{i=1}^{G} \frac{K_{i}}{K} d_{i}^2  \bar{e}_{i,g,j} \textbf{R}_{BS_{i}} \textbf{Q}_{i} \left( \textbf{Z}_{i} \textbf{Z}_{i}^{H} + \frac{1}{\rho_{tr}} \textbf{I}_{N} \right)\textbf{Q}^{H}_{i} \textbf{R}^{H}_{BS_{i}}}{ -  \frac{K_{g}}{K} d_{g}^2  \bar{e}_{g,g,j} \textbf{R}_{BS_{g}} \textbf{Q}_{g}  \textbf{z}_{g,j} \textbf{z}_{g,j}^{H} \textbf{Q}^{H}_{g} \textbf{R}^{H}_{BS_{g}} + \alpha \textbf{I}_{N} \Big)^{-1} \textbf{R}_{BS_{g}} \textbf{Q}_{g} \textbf{z}_{g,j}}}{ \splitfrac{1+ \bar{e}_{g,g,j} \frac{K_{g}}{K} d_{g}^2 \textbf{z}_{g,j}^{H} \textbf{Q}^{H}_{g} \textbf{R}^{H}_{BS_{g}} \Big(\sum_{i=1}^{G} \frac{K_{i}}{K} d_{i}^2  \bar{e}_{i,g,j} \textbf{R}_{BS_{i}} \textbf{Q}_{i} \left( \textbf{Z}_{i} \textbf{Z}_{i}^{H} + \frac{1}{\rho_{tr}} \textbf{I}_{N} \right)}{\times \textbf{Q}^{H}_{i} \textbf{R}^{H}_{BS_{i}} -  \frac{K_{g}}{K} d_{g}^2  \bar{e}_{g,g,j} \textbf{R}_{BS_{g}} \textbf{Q}_{g}  \textbf{z}_{g,j} \textbf{z}_{g,j}^{H} \textbf{Q}^{H}_{g} \textbf{R}^{H}_{BS_{g}} + \alpha \textbf{I}_{N} \Big)^{-1} \textbf{R}_{BS_{g}} \textbf{Q}_{g} \textbf{z}_{g,j}}} \right). \nonumber 
\end{align}
Then from trace lemma and rank-one perturbation lemma (Lemmas 3 and 5 in \cite{linearprecoding1}) we have,
\begin{align}
&T_{1}- \frac{d_{g}^2 }{K} \sum_{j=1}^{S_{g}} \frac{\frac{\bar{\textbf{s}}_{g,j}}{S_{g}} \text{tr } \bar{\textbf{D}}_{g} \left(\sum_{i=1}^{G}  \bar{e}_{i} \tilde{\boldsymbol{\Phi}}_{i} + \alpha \textbf{I}_{N} \right)^{-1}  }{1+\bar{e}_{g} \frac{K_{g}}{K} d_{g}^2 \frac{\bar{\textbf{s}}_{g,j}}{S_{g}} \text{tr } \bar{\textbf{D}}_{g} \left(\sum_{i=1}^{G}  \bar{e}_{i} \tilde{\boldsymbol{\Phi}}_{i} + \alpha \textbf{I}_{N} \right)^{-1} } \xrightarrow[N \rightarrow \infty]{a.s.} 0,
\end{align}
where $\bar{\textbf{D}}_{g}=\textbf{R}_{BS_{g}} \textbf{Q}_{g} \textbf{R}_{BS_{g}} \textbf{Q}^{H}_{g} \textbf{R}^{H}_{BS_{g}}$.

Notice that $\tilde{\boldsymbol{\Phi}}_{i}$ is still a function of $\textbf{Z}_{i}$ which is random. To remove the dependence of $T_{1}$ on $\textbf{Z}_{i}$, we need the deterministic equivalent of $\frac{1}{S_{g}} \text{tr }\bar{\textbf{D}}_{g} \left(\sum_{i=1}^{G}  \bar{e}_{i} \tilde{\boldsymbol{\Phi}}_{i} + \alpha \textbf{I}_{N} \right)^{-1}$. Using the expression of $\tilde{\boldsymbol{\Phi}}_{i}$ in (\ref{MMSE_estimatecorr}) and Corollary 1 from \cite{theorem}, we have the following convergence,
\begin{align}
\label{delta_ex}
&\frac{1}{S_{g}} \text{tr }\textbf{R}_{g} \left(\sum_{i=1}^{G}  \bar{e}_{i} \tilde{\boldsymbol{\Phi}}_{i} + \alpha \textbf{I}_{N} \right)^{-1} - \frac{1}{S_{g}} \text{tr }\textbf{R}_{g}  (\sum_{i=1}^{G} \bar{f}_{i} \bar{\textbf{D}}_{i} +\frac{K_{i}}{K} d_{i}^2 \frac{\bar{e}_{i}}{\rho_{tr}} \tilde{\textbf{D}}_{i} +\alpha \textbf{I}_{N})^{-1}   \xrightarrow[N \rightarrow \infty]{a.s.} 0,
\end{align}
where ${\textbf{R}}_{g}$ is any deterministic matrix that satisfies $\text{lim sup}_{N} ||\textbf{R}_{g}|| < \infty$ and,
\begin{align}
&\bar{f}_{g}=\frac{1}{S_{g}} \sum_{j=1}^{S_{g}} \frac{\frac{K_{g}}{K} d_{g}^2 \bar{e}_{g}\bar{s}_{g,j}}{1+\frac{K_{g}}{K} d_{g}^2 f_{g}(\bar{\textbf{D}}_{g}) \bar{e}_{g} \bar{s}_{g,j}}, \\
\label{f_ex}
&f_{g}({\textbf{R}}_{g})=\frac{1}{S_{g}} \text{tr } \textbf{R}_{g}(\sum_{i=1}^{G} \bar{f}_{i} \bar{\textbf{D}}_{i} +\frac{K_{i}}{K} d_{i}^2 \frac{\bar{e}_{i}}{\rho_{tr}} \tilde{\textbf{D}}_{i} +\alpha \textbf{I}_{N})^{-1},
\end{align}
such that $(f_{g}({\textbf{R}}_{g}), \bar{f}_{g}) \geq 0$. Substituting $\bar{f}_{g}$ in $f_{g}({\textbf{R}}_{g})$, we have,
\begin{align}
\label{Th1_1}
&f_{g}({\textbf{R}}_{g})=\frac{1}{S_{g}} \text{tr } \textbf{R}_{g} \left(\sum_{i=1}^{G} \frac{\bar{\textbf{D}}_{i}}{S_{i}} \left( \sum_{j=1}^{S_{i}} \frac{\frac{K_{i}}{K} d_{i}^2 \bar{e}_{i}\bar{s}_{i,j}}{1+\frac{K_{i}}{K} d_{i}^2 f_{i}(\bar{\textbf{D}}_{i}) \bar{e}_{i} \bar{s}_{i,j}} \right)  +\frac{K_{i}}{K} d_{i}^2 \frac{\bar{e}_{i}}{\rho_{tr}} \tilde{\textbf{D}}_{i} +\alpha \textbf{I}_{N}\right)^{-1}.
\end{align}
Consequently we get the following convergence for $T_{1}$ and $T_{2}$, 
\begin{align}
&T_{1} - \frac{1}{K} d_{g}^2 \sum_{j=1}^{S_{g}} \frac{\bar{s}_{g,j} f_{g}(\bar{\textbf{D}}_{g})} {1+ \frac{K_{g}}{K} \bar{e}_{g} d_{g}^2 \bar{s}_{g,j} f_{g}(\bar{\textbf{D}}_{g})}   \xrightarrow[N \rightarrow \infty]{a.s.} 0, \\
&T_{2} -\frac{S_{g}}{K} d_{g}^2 \frac{1}{\rho_{tr}} f_{g}(\tilde{\textbf{D}}_{g})   \xrightarrow[N \rightarrow \infty]{a.s.} 0.
\end{align}
Plugging $T_{1}$ and $T_{2}$ into the expression of $e_{g}$ in (\ref{T_ex}) yields,
\begin{align}
& e_{g}=\frac{1}{K} d_{g}^2 \left( \sum_{j=1}^{S_{g}} \frac{\bar{s}_{g,j} f_{g}(\bar{\textbf{D}}_{g})}{1+\frac{K_{g}}{K} d_{g}^{2} \bar{s}_{g,j} \bar{e}_{g} f_{g}(\bar{\textbf{D}}_{g})} + \frac{S_{g}}{\rho_{tr}} f_{g}(\tilde{\textbf{D}}_{g})\right)  + \epsilon_{g},  
\end{align}
where $\epsilon_{g} \xrightarrow[N \rightarrow \infty]{a.s.} 0$. Consider the deterministic counterpart of $(e_{g}(\alpha), \bar{e}_{g}(\alpha), f_{g}(\textbf{R}_{g},\alpha))$ as $(m_{g}(\alpha), \\ \bar{m}_{g}(\alpha), \delta_{g}(\textbf{R}_{g}, \alpha))$  defined in (\ref{fixedpoint}) and define $\Upsilon_{1}=\text{max}_{g} |e_{g}(\alpha)-m_{g}(\alpha)|$, $\Upsilon_{2}=\text{max }_{g}|\bar{e}_{g}(\alpha)-\bar{m}_{g}(\alpha)|$, $\Upsilon_{3}=\text{max}_{g}|f_{g}({\textbf{R}}_{g},\alpha)-\delta_{g}({\textbf{R}}_{g},\alpha)|$ and $\epsilon=\text{max}_{g} |\epsilon_{g}|$. It can be shown using the techniques from Appendix E of \cite{iterative} that,
\begin{align}
\label{con1}
&\Upsilon_{1}, \Upsilon_{2}, \Upsilon_{3} \xrightarrow[N \rightarrow \infty]{a.s.} 0,
\end{align} 
for $\alpha$ sufficiently large and $\epsilon \xrightarrow[N \rightarrow \infty]{a.s.} 0$. This result can be extended to all $\alpha$ by Vitali convergence theorem \cite{Vitali}.

Now combining (\ref{ref1}) with the result in (\ref{delta_ex}) for $\textbf{R}_{g}=\textbf{U}$ will yield,
\begin{align}
&\frac{1}{K}\textbf{U}\hat{\textbf{C}}^{-1}(\alpha)-\frac{1}{K}\textbf{U}\bar{\textbf{T}}(\alpha) \xrightarrow[N \rightarrow \infty]{a.s.} 0,
\end{align}
where $\bar{\textbf{T}}(\alpha)$ is given by (\ref{T_bar}). This completes the proof of Theorem 1. The uniqueness of the solution  $(m_{g}(\alpha), \bar{m}_{g}(\alpha), \delta_{g}(\textbf{R}_{g}, \alpha))$ can be proved by showing that the G-variate function in (\ref{fixedpoint}) is a standard interference function  as done for Theorem 2 in \cite{fixedpoint1}.


\vspace{-.13in}
\section{Proof of Lemma 2}
\vspace{-.06in}
\subsection{ Deterministic equivalent of $\frac{1}{K} \hat{\textbf{h}}_{k,g}^{H} \hat{\textbf{C}}_{[k,g]}^{-1} \hat{\textbf{h}}_{k,g}$:}

It has been shown in (\cite{iterative}, Appendix E) that $\textbf{Z}_{g}$ in (\ref{DS1}) satisfies $\text{lim sup}_{N}||\textbf{Z}_{g} \textbf{Z}_{g}^{H}||< \infty$ almost surely. We can therefore assume $\textbf{Z}_{g}$ to be deterministic as a starting point and use trace lemma and rank-one lemma (Lemma 3 and Lemma 5 from \cite{linearprecoding1}) to have,
\begin{align}
\label{use1}
&\frac{1}{K} \hat{\textbf{h}}_{k,g}^{H} \hat{\textbf{C}}_{[k,g]}^{-1} \hat{\textbf{h}}_{k,g}-  \frac{1}{K} \text{tr }  \boldsymbol{\Phi}_{g} \hat{\textbf{C}}^{-1}  \xrightarrow[N \rightarrow \infty]{a.s.} 0,
\end{align}
where $\boldsymbol{\Phi}_{g}$ is given by (\ref{MMSE_estimatecorr}). Utilizing the result in (\ref{ref1}) for $\textbf{U}=\boldsymbol{\Phi}_{g}$ yields $\frac{1}{K}\text{tr } \boldsymbol{\Phi}_{g} \hat{\textbf{C}}^{-1} - e_{g} \xrightarrow[N \rightarrow \infty]{a.s.} 0$. The result can be extended  to random $\textbf{Z}_{g}$s using Fubini Theorem and the deterministic equivalent of $e_{g}$ is obtained as $m_{g}$ in Appendix B.
\vspace{-.11in}
\subsection{ Deterministic equivalent of $\frac{1}{K} {\textbf{h}}_{k,g}^{H} \hat{\textbf{C}}_{[k,g]}^{-1} \hat{\textbf{h}}_{k,g}$:}

Utilizing the expressions of the double scattering channel in (\ref{DS1}) and its MMSE estimate in (\ref{MMSE_estimate}) and the independence of $\tilde{\textbf{w}}_{k,g}$ and $\textbf{n}_{k,g}^{tr}$, we have,
\begin{align}
&\frac{1}{K} {\textbf{h}}_{k,g}^{H} \hat{\textbf{C}}_{[k,g]}^{-1} \hat{\textbf{h}}_{k,g}- \frac{S_{g} d_{g}}{K} \tilde{\textbf{w}}_{k,g}^{H} \textbf{Z}^{H}_{g} \hat{\textbf{C}}_{[k,g]}^{-1} \textbf{R}_{BS_{g}} \textbf{Q}_{g} \textbf{Z}_{g} \tilde{\textbf{w}}_{k,g}   \xrightarrow[N \rightarrow \infty]{a.s.} 0.
\end{align}

 Next exploit trace lemma and rank-1 lemma (Lemmas 3 and 5 from \cite{linearprecoding1}) to get,
\begin{align}
&\frac{1}{K} {\textbf{h}}_{k,g}^{H} \hat{\textbf{C}}_{[k,g]}^{-1} \hat{\textbf{h}}_{k,g}- \frac{1}{K_{g}} \text{tr }\tilde{ \bar{\boldsymbol{\Phi}}}_{g} \hat{\textbf{C}}^{-1}  \xrightarrow[N \rightarrow \infty]{a.s.} 0,
\end{align}
where $\hat{\textbf{C}}$ is given by (\ref{C}) under the assumption of deterministic $\textbf{Z}_{g}$ and $\tilde{\bar{\boldsymbol{\Phi}}}_{g}=\frac{K_{g}}{K}\bar{\boldsymbol{\Phi}}_{g}$, where $\bar{\boldsymbol{\Phi}}_{g}= d_{g} \textbf{R}_{BS_{g}} \textbf{Q}_{g} \textbf{Z}_{g} \textbf{Z}_{g}^{H}$. Utilizing the result in (\ref{ref1}) for $\textbf{U}=\bar{\boldsymbol{\Phi}}_{g}$, we have,
\begin{align}
\label{ref2}
&\frac{1}{K_{g}} \text{tr }\tilde{ \bar{\boldsymbol{\Phi}}}_{g} \hat{\textbf{C}}^{-1} - \frac{1}{K_{g}} \text{tr }\tilde{ \bar{\boldsymbol{\Phi}}}_{g}  (\sum_{i=1}^{G} \bar{e}_{i} \tilde{ \boldsymbol{\Phi}}_{i}  + \alpha \textbf{I}_{N})^{-1} \xrightarrow[N \rightarrow \infty]{a.s.} 0,
\end{align}
where $(e_{g}, \bar{e}_{g})$ are given as a unique solution to the set of implicit equations in (\ref{e_quant}), (\ref{e_quant1}), Denote $\frac{1}{K_{g}} \text{tr }\tilde{ \bar{\boldsymbol{\Phi}}}_{g}  (\sum_{i=1}^{G} \bar{e}_{i} \tilde{ \boldsymbol{\Phi}}_{i}  + \alpha \textbf{I}_{N})^{-1}$ in (\ref{ref2}) as $T_{3}$ and let $\textbf{Z}_{g}$'s be random now. A similar development as done for the term $T_1$ in Appendix B will yield,
\begin{align}
&T_{3} - \frac{ d_{g}}{K} \sum_{j=1}^{S_{g}} \frac{\frac{\bar{s}_{g,j}}{S_{g}} \text{tr }  \textbf{R}_{BS_{g}} \left(\sum_{i=1}^{G}  \bar{e}_{i} \tilde{\boldsymbol{\Phi}}_{i}+ \alpha \textbf{I}_{N} \right)^{-1} \textbf{R}_{BS_{g}} \textbf{Q}_{g} }{ 1+ \bar{e}_{g} \frac{K_{g}}{K} d_{g}^2  \frac{\bar{s}_{g,j}}{S_{g}} \text{tr } \bar{\textbf{D}}_{g} \left(\sum_{i=1}^{G}  \bar{e}_{i}\tilde{\boldsymbol{\Phi}}_{i} + \alpha \textbf{I}_{N} \right)^{-1}}  \xrightarrow[N \rightarrow \infty]{a.s.} 0.
\end{align}
Using the convergence in (\ref{delta_ex}) and $f_{g}(\textbf{R}_{g})$ defined in (\ref{f_ex}), we obtain,
\begin{align}
\label{T_3}
&T_{3}- \frac{1}{K} d_{g} \sum_{j=1}^{S_{g}} \frac{\bar{s}_{g,j} f_{g}(\textbf{R}_{BS_{g}}\textbf{Q}_{g} \textbf{R}_{BS_{g}}) }{ 1+ \bar{e}_{g} \frac{K_{g}}{K} d_{g}^2  \bar{s}_{g,j} f_{g}(\bar{\textbf{D}}_{g}) }  \xrightarrow[N \rightarrow \infty]{a.s.} 0.
\end{align}
Replace $\bar{e}_{g}$ and $f_{g}(\textbf{R}_{g})$ in (\ref{T_3}) by their deterministic approximations $\bar{m}_{g}$ and $\delta_{g}(\textbf{R}_{g})$ respectively from Appendix B. This yields the deterministic equivalent of $\frac{1}{K} \textbf{h}_{k,g}^{H} \hat{\textbf{C}}_{[k,g]}^{-1} \hat{\textbf{h}}_{k,g}$ denoted as $h_{g}$.

\vspace{-.15in}
\section{Proof of Theorem 2}

\vspace{-.1in}
We are interested in finding the deterministic equivalent of:
\begin{align}
\label{chi_quant}
&\chi_{g}=\frac{1}{K^2} \text{tr } \boldsymbol{\Phi}_{g} \hat{\textbf{C}}^{-1} \hat{\textbf{H}}^{H} \textbf{P} \hat{\textbf{H}} \hat{\textbf{C}}^{-1}.
\end{align} 

First we need to control the variance of ${\chi}_{g}$ and prove that $\text{var}(\chi_{g})$ converges to zero. This can be done using standard tools from RMT (see  \cite{walid_mutual_information}) and will imply that,
\begin{align}
&\chi_{g}- \chi^{o}_{g} \xrightarrow[N \rightarrow \infty]{a.s.} 0,
\end{align}
where $\chi^{o}_{g} =\mathbb{E}[\chi_{g}]$. This allows us to focus directly on $\mathbb{E}[\chi_{g}]$.

Using the expression of $\boldsymbol{\Phi}_{g}$ in (\ref{MMSE_estimatecorr}) we have,
\begin{align}
\label{chi_1}
&\chi^{o}_{g}=\frac{1}{K^2} d_{g}^2 \mathbb{E}\left[\text{tr } \textbf{R}_{BS_{g}} \textbf{Q}_{g} \left( \textbf{Z}_{g} \textbf{Z}_{g}^{H} + \frac{1}{\rho_{tr}} \textbf{I}_{N} \right)\textbf{Q}^{H}_{g} \textbf{R}^{H}_{BS_{g}} \hat{\textbf{C}}^{-1} \hat{\textbf{H}}^{H} \textbf{P} \hat{\textbf{H}} \hat{\textbf{C}}^{-1}\right], \\
\label{chi_11}
&=d_{g}^2 \left(\kappa^{o}_{g}(\textbf{R}_{BS_{g}} \textbf{Q}_{g},\textbf{Q}^{H}_{g} \textbf{R}^{H}_{BS_{g}} ) + \frac{1}{\rho_{tr}}\xi^{o}(\textbf{R}_{BS_{g}} \textbf{Q}_{g},\textbf{Q}^{H}_{g} \textbf{R}^{H}_{BS_{g}})\right),
\end{align}
where,
\vspace{-.06in}
\begin{align}
&\kappa_{g}(\textbf{R}_{BS_{g}} \textbf{Q}_{g}, \textbf{Q}^{H}_{g} \textbf{R}^{H}_{BS_{g}} )=\frac{1}{K^2} \text{tr } \textbf{R}_{BS_{g}} \textbf{Q}_{g} \textbf{Z}_{g} \textbf{Z}_{g}^{H} \textbf{Q}^{H}_{g} \textbf{R}^{H}_{BS_{g}} \hat{\textbf{C}}^{-1} \hat{\textbf{H}}^{H} \textbf{P} \hat{\textbf{H}} \hat{\textbf{C}}^{-1}, \\
&\xi(\textbf{R}_{BS_{g}} \textbf{Q}_{g}, \textbf{Q}^{H}_{g} \textbf{R}^{H}_{BS_{g}} )=\frac{1}{K^2} \text{tr } \textbf{R}_{BS_{g}} \textbf{Q}_{g} \textbf{Q}^{H}_{g} \textbf{R}^{H}_{BS_{g}} \hat{\textbf{C}}^{-1} \hat{\textbf{H}}^{H} \textbf{P} \hat{\textbf{H}} \hat{\textbf{C}}^{-1},
\end{align}
and $\kappa^{o}_{g}$ and $\xi^{o}$ denote their respective expectations. We therefore start the proof by deriving the deterministic equivalents of $\kappa_{g}(\textbf{A},\textbf{B})$ and $\xi(\textbf{A},\textbf{B})$, where \textbf{A} and \textbf{B} are  deterministic matrices of uniformly bounded spectral norm. Plugging these results in (\ref{chi_11}) will complete the proof of Theorem 2.

\vspace{-.1in}
\subsection{ Deterministic equivalent of $\kappa_{g}(\textbf{A},\textbf{B})=\frac{1}{K^2} \text{tr } \textbf{A} \textbf{Z}_{g} \textbf{Z}_{g}^{H} \textbf{B} \hat{\textbf{C}}^{-1} \hat{\textbf{H}}^{H} \textbf{P} \hat{\textbf{H}} \hat{\textbf{C}}^{-1}$:}

The aim of this section is to derive a deterministic equivalent for the random quantity $\kappa_{g}(\textbf{A},\textbf{B})$. We can again control the variance of $\kappa_{g}$ and show that it goes to zero implying that,
\vspace{-.06in}
\begin{align}
\label{control}
&\kappa_{g}(\textbf{A},\textbf{B})- \kappa^{o}_{g}(\textbf{A},\textbf{B}) \xrightarrow[N \rightarrow \infty]{a.s.} 0,
\end{align}
where $\kappa^{o}_{g}(\textbf{A},\textbf{B})=\mathbb{E}[\kappa_{g}(\textbf{A},\textbf{B})]$. Using the result from last section in (\ref{ref1}) that $\frac{1}{K}\text{tr }\hat{\textbf{C}}^{-1}-\frac{1}{K}\text{tr }\textbf{T} \xrightarrow[N \rightarrow \infty]{a.s.} 0$, where $\textbf{T}=(\sum_{i=1}^{G} \frac{\tilde{\boldsymbol{\Phi}}_{i}}{1+m_{i}} +\alpha \textbf{I}_{N})^{-1}$ and the resolvent identity given as $\hat{\textbf{C}}^{-1}-\textbf{T}=\textbf{T}(\textbf{T}^{-1}-\hat{\textbf{C}})\hat{\textbf{C}}^{-1}=\textbf{T}\left( \sum_{i=1}^{G} \frac{\tilde{\boldsymbol{\Phi}}_{i}}{1+m_{i}} - \frac{1}{K}\hat{\textbf{H}}_{i}^{H} \hat{\textbf{H}}_{i} \right) \hat{\textbf{C}}^{-1}$, we decompose $\kappa_{g}$ as,
\begin{align}
&\kappa_{g}(\textbf{A},\textbf{B})=\frac{1}{K^2} \text{tr } \textbf{A} \textbf{Z}_{g} \textbf{Z}_{g}^{H} \textbf{B} \textbf{T} \hat{\textbf{H}}^{H} \textbf{P} \hat{\textbf{H}} \hat{\textbf{C}}^{-1} + \frac{1}{K^2} \sum_{i=1}^{G}  \text{tr } \textbf{A} \textbf{Z}_{g} \textbf{Z}_{g}^{H}\textbf{B} \textbf{T} \frac{\tilde{\boldsymbol{\Phi}}_{i}}{1+m_{i}} \hat{\textbf{C}}^{-1} \hat{\textbf{H}}^{H} \textbf{P} \hat{\textbf{H}} \hat{\textbf{C}}^{-1} \nonumber \\
&- \frac{1}{K^3} \sum_{i=1}^{G}  \text{tr } \textbf{A} \textbf{Z}_{g} \textbf{Z}_{g}^{H}\textbf{B} \textbf{T}  \hat{\textbf{H}}_{i}^{H} \hat{\textbf{H}}_{i} \hat{\textbf{C}}^{-1} \hat{\textbf{H}}^{H} \textbf{P} \hat{\textbf{H}} \hat{\textbf{C}}^{-1}= Z_{1}+Z_{2}+Z_{3}.
\end{align}
We start by applying matrix inversion lemma (Lemma 1 in \cite{massiveMIMOO}) on $\mathbb{E}[Z_{1}]$ and re-writing the expression to get,
\begin{align}
&\mathbb{E}[Z_{1}]=\frac{1}{K^2} \sum_{i=1}^{G}\sum_{l=1}^{K_{i}} p_{l,i} \Big(\mathbb{E} \left[ \frac{\hat{\textbf{h}}_{l,i}^{H} \hat{\textbf{C}}_{[l,i]}^{-1} \textbf{A} \textbf{Z}_{g} \textbf{Z}_{g}^{H}\textbf{B} \textbf{T} \hat{\textbf{h}}_{l,i}\left(\frac{1}{K} \text{tr } \boldsymbol{\Phi}_{i} \hat{\textbf{C}}^{-1}_{[l,i]} - \frac{1}{K} \hat{\textbf{h}}_{l,i}^{H} \hat{\textbf{C}}_{[l,i]}^{-1} \hat{\textbf{h}}_{l,i}\right)}{(1+\frac{1}{K}\hat{\textbf{h}}_{l,i}^{H} \hat{\textbf{C}}_{[l,i]}^{-1} \hat{\textbf{h}}_{l,i})(1+\frac{1}{K} \text{tr } \boldsymbol{\Phi}_{i} \hat{\textbf{C}}^{-1}_{[l,i]})} \right]\nonumber \\
&+\mathbb{E}\left[\frac{\hat{\textbf{h}}_{l,i}^{H} \hat{\textbf{C}}_{[l,i]}^{-1} \textbf{A} \textbf{Z}_{g} \textbf{Z}_{g}^{H} \textbf{B} \textbf{T} \hat{\textbf{h}}_{l,i}}{1+\frac{1}{K} \text{tr } \boldsymbol{\Phi}_{i} \hat{\textbf{C}}^{-1}_{[l,i]}} \right]   \Big).
\end{align} 
Assuming $\textbf{Z}_{g}$ to be deterministic and using the fact that $\text{lim sup}_{N} ||\textbf{Z}_{g}\textbf{Z}_{g}^{H}|| < \infty$ almost surely, we can use the expression of $\hat{\textbf{h}}_{k,g}$ in (\ref{MMSE_estimate1}) and show with the help of trace lemma (Lemma 3 from \cite{linearprecoding1}) that the first term on the right side of the above equation is negligible. Now using the rank-one perturbation lemma (Lemma 5 from \cite{linearprecoding1}) and Lemma 2 on the second term yields,
\begin{align}
&\mathbb{E}[Z_{1}]=\frac{1}{K^2} \sum_{i=1}^{G}\sum_{l=1}^{K_{i}} p_{l,i} \frac{\mathbb{E}\left[\hat{\textbf{h}}_{l,i}^{H} \hat{\textbf{C}}_{[l,i]}^{-1} \textbf{A} \textbf{Z}_{g} \textbf{Z}_{g}^{H} \textbf{B} \textbf{T} \hat{\textbf{h}}_{l,i}\right]}{1+m_{i}}  + o(1).
\end{align}

Using trace lemma (Lemma 3 from \cite{linearprecoding1}) with the expression of $\hat{\textbf{h}}_{k,g}$ in (\ref{MMSE_estimate1}) we have,
\begin{align}
&\mathbb{E}[Z_{1}]=\frac{1}{K} \sum_{i=1}^{G}\sum_{l=1}^{K_{i}} \frac{p_{l,i}}{K_{i}} \frac{\mathbb{E}\left[ \text{tr } \tilde{\boldsymbol{\Phi}}_{i} \hat{\textbf{C}}_{[l,i]}^{-1} \textbf{A} \textbf{Z}_{g} \textbf{Z}_{g}^{H} \textbf{B} \textbf{T} \right]}{1+m_{i}}  + o(1).
\end{align}

Now using the rank-one lemma (Lemma 5 from \cite{linearprecoding1}) and Corollary 1 from \cite{theorem} we have,
\begin{align}
&\mathbb{E}[Z_{1}]=\frac{1}{K} \sum_{i=1}^{G}\sum_{l=1}^{K_{i}} \frac{p_{l,i}}{K_{i}} \frac{\mathbb{E}\left[ \text{tr } \tilde{\boldsymbol{\Phi}}_{i} \textbf{T} \textbf{A} \textbf{Z}_{g} \textbf{Z}_{g}^{H} \textbf{B} \textbf{T} \right]}{1+m_{i}}  + o(1).
\end{align}

Note that the analysis in the last two steps can be extended to random $\textbf{Z}_{g}$s using Fubini Theorem and we need the deterministic equivalent of $\bar{\beta}_{g,i}(\textbf{A},\textbf{B})=\frac{1}{K_{i}}\mathbb{E}[\text{tr } \tilde{\boldsymbol{\Phi}}_{i} \textbf{T} \textbf{A} \textbf{Z}_{g} \textbf{Z}_{g}^{H} \textbf{B} \textbf{T}]$ under the actual random $\textbf{Z}_{g}$s. Using the expression of $\boldsymbol{\Phi}_{i}$ from (\ref{MMSE_estimatecorr}) results in $\bar{\beta}_{g,i}(\textbf{A},\textbf{B})=$
\begin{align}
& \mathbb{E}\left[\frac{d_{i}^2}{K} \sum_{j=1}^{S_{i}} \textbf{z}_{i,j}^{H} \textbf{Q}^{H}_{i} \textbf{R}^{H}_{BS_{i}} \textbf{T} \textbf{A} \textbf{Z}_{g} \textbf{Z}_{g}^{H} \textbf{B} \textbf{T} \textbf{R}_{BS_{i}} \textbf{Q}_{i} \textbf{z}_{i,j}\right] \hspace{-.02in} + \hspace{-.02in} \mathbb{E}\left[\frac{d_{i}^2}{K \rho_{tr}} \text{tr } \textbf{R}_{BS_{i}} \textbf{Q}_{i} \textbf{Q}^{H}_{i} \textbf{R}^{H}_{BS_{i}} \textbf{T} \textbf{A} \textbf{Z}_{g} \textbf{Z}_{g}^{H}\textbf{B} \textbf{T} \right] \nonumber\\
&\hspace{-.03in}=  T_{4}+T_{5}.
\end{align}
We analyze the two terms separately. First we use matrix inversion lemma (Lemma 1 from \cite{massiveMIMOO}) on $T_{4}$ to remove the dependence of $\textbf{T}$ on $\textbf{z}_{i,j}$ resulting in $T_{4}=$
\begin{align}
&\hspace{-.05in} \begin{cases} \hspace{-.05in} \frac{d_{i}^2}{K} \sum_{j=1}^{S_{i}} \mathbb{E}\left[ \frac{\textbf{z}_{i,j}^{H} \textbf{Q}^{H}_{i} \textbf{R}^{H}_{BS_{i}} \tilde{\textbf{T}}_{i} \textbf{A} \textbf{Z}_{g} \textbf{Z}_{g}^{H} \textbf{B} \tilde{\textbf{T}}_{i} \textbf{R}_{BS_{i}} \textbf{Q}_{i} \textbf{z}_{i,j}}{(1+\bar{m}_{i}\frac{K_{i}}{K} d_{i}^2 \textbf{z}_{i,j}^{H} \textbf{Q}^{H}_{i} \textbf{R}^{H}_{BS_{i}} \tilde{\textbf{T}}_{i}  \textbf{R}_{BS_{i}} \textbf{Q}_{i} \textbf{z}_{i,j} )^2} \right] & \hspace{-.12in} \text{if } g\neq i, \\ 
\hspace{-.05in} \frac{d_{i}^2}{K} \sum_{j=1}^{S_{i}} \mathbb{E}\left[ \frac{\textbf{z}_{i,j}^{H} \textbf{Q}^{H}_{i} \textbf{R}^{H}_{BS_{i}} \tilde{\textbf{T}}_{i} \textbf{A} (\textbf{Z}_{g} \textbf{Z}_{g}^{H} -\textbf{z}_{i,j}\textbf{z}_{i,j}^{H}) \textbf{B} \tilde{\textbf{T}}_{i}  \textbf{R}_{BS_{i}} \textbf{Q}_{i} \textbf{z}_{i,j}+\textbf{z}_{i,j}^{H} \textbf{B}  \tilde{\textbf{T}}_{i} \textbf{R}_{BS_{i}} \textbf{Q}_{i} \textbf{z}_{i,j} \textbf{z}_{i,j}^{H} \textbf{Q}_{i}^{H} \textbf{R}_{BS_{i}}^{H} \tilde{\textbf{T}}_{i} \textbf{A} \textbf{z}_{i,j}}{(1+\bar{m}_{i}\frac{K_{i}}{K} d_{i}^2 \textbf{z}_{i,j}^{H} \textbf{Q}^{H}_{i} \textbf{R}^{H}_{BS_{i}} \tilde{\textbf{T}}_{i}  \textbf{R}_{BS_{i}} \textbf{Q}_{i} \textbf{z}_{i,j} )^2} \right] & \hspace{-.12in}\text{if } g=i,
\end{cases} \nonumber
\end{align} 
where $\tilde{\textbf{T}}_{i}=\Big(\sum_{l=1}^{G} \frac{K_{l}}{K} d_{l}^2  \bar{m}_{l} \textbf{R}_{BS_{l}} \textbf{Q}_{l} \left( \textbf{Z}_{l} \textbf{Z}_{l}^{H} + \frac{1}{\rho_{tr}} \textbf{I}_{N} \right)\textbf{Q}^{H}_{l} \textbf{R}^{H}_{BS_{l}} -  \frac{K_{i}}{K} d_{i}^2  \bar{m}_{i} \textbf{R}_{BS_{i}} \textbf{Q}_{i}  \textbf{z}_{i,j} \textbf{z}_{i,j}^{H} \textbf{Q}^{H}_{i} \textbf{R}^{H}_{BS_{i}} \\ + \alpha \textbf{I}_{N} \Big)^{-1}$. Next we use trace lemma and rank-one lemma (Lemmas 3 and 5 from \cite{linearprecoding1}) as, 
\begin{align}
&T_{4}= \begin{cases}\frac{d_{i}^2}{K} \sum_{j=1}^{S_{i}} \mathbb{E}\left[\frac{\frac{\bar{s}_{i,j}}{S_{i}} \text{tr } \textbf{R}_{BS_{i}} \textbf{Q}^{H}_{i} \textbf{R}^{H}_{BS_{i}} \textbf{T} \textbf{A} \textbf{Z}_{g} \textbf{Z}_{g}^{H}\textbf{B} \textbf{T} \textbf{R}_{BS_{i}} \textbf{Q}_{i} }{(1+\bar{m}_{i}\frac{K_{i}}{K} d_{i}^2 \frac{\bar{s}_{i,j}}{S_{i}} \text{tr } \textbf{R}_{BS_{i}} \textbf{Q}^{H}_{i} \textbf{R}^{H}_{BS_{i}} \textbf{T}  \textbf{R}_{BS_{i}} \textbf{Q}_{i})^2 } \right] + o(1) & \hspace{-.06in} \text{if } g\neq i, \\ 
 \frac{d_{i}^2}{K} \sum_{j=1}^{S_{i}} \mathbb{E}\left[\frac{\splitfrac{\frac{\bar{s}_{i,j}}{S_{i}} \text{tr } \textbf{R}_{BS_{i}} \textbf{Q}^{H}_{i} \textbf{R}^{H}_{BS_{i}} \textbf{T} \textbf{A} (\textbf{Z}_{g} \textbf{Z}_{g}^{H} -\textbf{z}_{i,j}\textbf{z}_{i,j}^{H}) \textbf{B} \textbf{T}  \textbf{R}_{BS_{i}} \textbf{Q}_{i}}{ + (\frac{\bar{s}_{i,j}}{S_{i}} \text{tr } \textbf{R}_{BS_{i}} \textbf{B}  \textbf{T} \textbf{R}_{BS_{i}} \textbf{Q}_{i}) (\frac{\bar{s}_{i,j}}{S_{i}} \text{tr } \textbf{R}_{BS_{i}} \textbf{Q}_{i}^{H} \textbf{R}_{BS_{i}}^{H} \textbf{T} \textbf{A} )}}{(1+\bar{m}_{i}\frac{K_{i}}{K} d_{i}^2 \frac{\bar{s}_{i,j}}{S_{i}} \text{tr } \textbf{R}_{BS_{i}} \textbf{Q}^{H}_{i} \textbf{R}^{H}_{BS_{i}} \textbf{T}  \textbf{R}_{BS_{i}} \textbf{Q}_{i})^2  } \right] + o(1) & \hspace{-.06in} \text{if } g=i. \nonumber
\end{cases}
\end{align}

Using matrix inversion lemma to remove the dependence of \textbf{T} on $\textbf{z}_{g,j}$ and using $\frac{1}{S_{i}} \text{tr }\bar{\textbf{D}}_{i} \textbf{T} - \delta_{i}(\bar{\textbf{D}}_{i})\xrightarrow[N \rightarrow \infty]{a.s.} 0$ from Appendix B, where $\bar{\textbf{D}}_{i}= \textbf{R}_{BS_{i}} \textbf{Q}_{i} \textbf{R}_{BS_{i}}  \textbf{Q}^{H}_{i} \textbf{R}^{H}_{BS_{i}}$, yields,
\begin{align}
&T_{4} = \begin{cases}\frac{d_{i}^2}{K} \sum_{j=1}^{S_{i}} \mathbb{E}\left[\frac{\bar{s}_{i,j}}{S_{i}} \sum_{n=1}^{S_{g}} \frac{\textbf{z}_{g,n}^{H} \textbf{B} \tilde{\textbf{T}}_{g} \textbf{R}_{BS_{i}} \textbf{Q}_{i} \textbf{R}_{BS_{i}}  \textbf{Q}^{H}_{i} \textbf{R}^{H}_{BS_{i}} \tilde{\textbf{T}}_{g} \textbf{A} \textbf{z}_{g,n}}{(1+\frac{K_{i}}{K} d_{i}^2 \bar{m}_{i} \bar{s}_{i,j} \delta_{i}(\bar{\textbf{D}}_{i}))^2(1+\bar{m}_{g}\frac{K_{g}}{K} d_{g}^2 \textbf{z}_{g,n}^{H} \textbf{Q}^{H}_{g} \textbf{R}^{H}_{BS_{g}} \tilde{\textbf{T}}_{g}  \textbf{R}_{BS_{g}} \textbf{Q}_{g} \textbf{z}_{g,n} )^2 } \right] & \text{if } g\neq i, \\ 
 \frac{d_{i}^2}{K} \sum_{j=1}^{S_{i}} \mathbb{E} \left[ \frac{\bar{s}_{i,j}}{S_{i}} \underset{n\neq j}{\sum_{n=1}^{S_{g}}} \frac{\textbf{z}_{g,n}^{H} \textbf{B} \tilde{\textbf{T}}_{g} \textbf{R}_{BS_{i}} \textbf{Q}_{i} \textbf{R}_{BS_{i}}  \textbf{Q}^{H}_{i} \textbf{R}^{H}_{BS_{i}} \tilde{\textbf{T}}_{g} \textbf{A} \textbf{z}_{g,n}}{(1+\frac{K_{i}}{K} d_{i}^2 \bar{m}_{i} \bar{s}_{i,j} \delta_{i}(\bar{\textbf{D}}_{i}))^2 (1+\bar{m}_{g}\frac{K_{g}}{K} d_{g}^2 \textbf{z}_{g,n}^{H} \textbf{Q}^{H}_{g} \textbf{R}^{H}_{BS_{g}} \tilde{\textbf{T}}_{g}  \textbf{R}_{BS_{g}} \textbf{Q}_{g} \textbf{z}_{g,n} )^2 } \right] \\
+ \frac{d_{i}^2}{K} \sum_{j=1}^{S_{i}} \mathbb{E}\left[\frac{\bar{s}_{i,j}^2 \delta_{i}(\textbf{R}_{BS_{i}} \textbf{Q}_{i} \textbf{R}_{BS_{i}} \textbf{B}) \delta_{i}(\textbf{A} \textbf{R}_{BS_{i}} \textbf{Q}_{i}^{H} \textbf{R}_{BS_{i}}^{H})} {(1+\frac{K_{i}}{K} d_{i}^2 \bar{m}_{i} \bar{s}_{i,j} \delta_{i}(\bar{\textbf{D}}_{i}))^2} \right] & \text{if } g=i. \nonumber
\end{cases}
\end{align}

Finally we use trace lemma and rank-one perturbation lemma from \cite{linearprecoding1} to get,
\begin{align}
&T_{4} = \begin{cases}\frac{d_{i}^2}{K} \sum_{j=1}^{S_{i}} \frac{\bar{s}_{i,j}}{S_{i}} \sum_{n=1}^{S_{g}}  \frac{ \bar{s}_{g,n} \mathbb{E}[\frac{1}{S_{g}}\text{tr }  \textbf{R}_{BS_{g}} \textbf{B} \textbf{T} \bar{\textbf{D}}_{i} \textbf{T} \textbf{A}] }{(1+\frac{K_{i}}{K} d_{i}^2 \bar{m}_{i} \bar{s}_{i,j} \delta_{i}(\bar{\textbf{D}}_{i}))^2(1+\frac{K_{g}}{K} d_{g}^2 \bar{m}_{g} \bar{s}_{g,n} \delta_{g}(\bar{\textbf{D}}_{g}))^2 } + o(1)& \text{if } g\neq i, \\ 
 \frac{d_{i}^2}{K} \sum_{j=1}^{S_{i}} \Big(\frac{\bar{s}_{i,j}}{S_{i}} \underset{n\neq j}{\sum_{n=1}^{S_{g}}}  \frac{ \bar{s}_{g,n}\mathbb{E}[ \frac{1}{S_{g}}\text{tr }  \textbf{R}_{BS_{g}} \textbf{B} \textbf{T} \bar{\textbf{D}}_{i} \textbf{T} \textbf{A} ]}{(1+\frac{K_{i}}{K} d_{i}^2 \bar{m}_{i} \bar{s}_{i,j} \delta_{i}(\bar{\textbf{D}}_{i}))^2 (1+\frac{K_{g}}{K} d_{g}^2 \bar{m}_{g} \bar{s}_{g,n} \delta_{g}(\bar{\textbf{D}}_{g}))^2 } \\ +  \frac{\bar{s}_{i,j}^2 \delta_{i}(\textbf{R}_{BS_{i}} \textbf{Q}_{i} \textbf{R}_{BS_{i}} \textbf{B}) \delta_{i}(\textbf{A} \textbf{R}_{BS_{i}} \textbf{Q}_{i}^{H} \textbf{R}_{BS_{i}}^{H})} {(1+\frac{K_{i}}{K} d_{i}^2 \bar{m}_{i} \bar{s}_{i,j} \delta_{i}(\bar{\textbf{D}}_{i}))^2}\Big) +o(1)& \text{if } g=i.\nonumber
\end{cases}
\end{align}

Similar steps would yield the following deterministic equivalent for $T_{5}$,
\begin{align}
&T_{5} = \frac{d_{i}^2}{K \rho_{tr}} \sum_{j=1}^{S_{g}} \frac{ \bar{s}_{g,j} \mathbb{E}[\frac{1}{S_{g}}\text{tr } \textbf{R}_{BS_{g}} \textbf{B} \textbf{T} \tilde{\textbf{D}}_{i} \textbf{T} \textbf{A}]}{(1+ \frac{K_{g}}{K} \bar{m}_{g} d_{g}^2 \bar{s}_{g,j} \delta_{g}(\bar{\textbf{D}}_{g}))^2} +o(1),
\end{align}
where $\tilde{\textbf{D}}_{i}=\textbf{R}_{BS_{i}} \textbf{Q}_{i}  \textbf{Q}^{H}_{i} \textbf{R}^{H}_{BS_{i}}$.

In order to complete the calculation of the deterministic equivalent of $\bar{\beta}_{g,i}(\textbf{A},\textbf{B})$, we need the deterministic equivalent of $u'_{g}(\textbf{R}_{g}, \textbf{L})=\mathbb{E}[\frac{1}{S_{g}}\text{tr } \textbf{R}_{g} \textbf{T} \textbf{L} \textbf{T}]$ which is stated in the following Lemma.

\textit{Lemma 3:} Define $u'_{g}(\textbf{R}_{g}, \textbf{L})=\mathbb{E}[\frac{1}{S_{g}}\text{tr } \textbf{R}_{g} \textbf{T} \textbf{L} \textbf{T}]$ where $\textbf{R}_{g}$ and $\textbf{L}$, $g= 1, \dots, G$ are deterministic matrices with uniformly bounded spectral norm. Then under the setting of assumptions \textbf{A-1}, \textbf{A-2} and \textbf{A-3} and for $\alpha>0$,
\begin{align}
&u'_{g}(\textbf{R}_{g}, \textbf{L})=\frac{1}{S_{g}} \text{tr } \textbf{R}_{g} \left[\bar{\textbf{T}} \left(\sum_{z=1}^{G} \frac{\bar{\textbf{D}}_{z} \bar{m}_{z}^2 }{S_{z}} \left(\frac{K_{z}}{K}\right)^2 d_{z}^4  \text{tr } (\bar{\textbf{S}}_{z} \textbf{W}_{z}^2 \bar{\textbf{S}}_{z}) u'_{z}(\bar{\textbf{D}}_{z}, \textbf{L})  + \textbf{L} \right) \bar{\textbf{T}}  \right] + o(1),
\end{align} 
where $\textbf{W}_{i}=\left( \textbf{I}_{S_{i}} + \frac{K_{i}}{K} d_{i}^2 \bar{m}_{i} \delta_{i}(\bar{\textbf{D}}_{i}) \bar{\textbf{S}}_{i} \right)^{-1}$ and $\textbf{u}'(\bar{\textbf{D}}, \textbf{L})=[u'_{1}(\bar{\textbf{D}}_{1}, \textbf{L}), u'_{2}(\bar{\textbf{D}}_{2}, \textbf{L}),\dots, u'_{G}(\bar{\textbf{D}}_{G}, \textbf{L}) ]^{T}$, which can be expressed as a system of linear equations as follows:
\begin{align}
\label{M_der_eq1}
&\textbf{u}'(\bar{\textbf{D}}, \textbf{L})=(\textbf{I}_{N}-\textbf{J}(\bar{\textbf{D}}))^{-1} \textbf{v}(\bar{\textbf{D}}, \textbf{L}), \\
&[\textbf{J}(\bar{\textbf{D}})]_{g,i}=\frac{1}{S_{g}} \text{tr }(\bar{\textbf{D}}_{g} \bar{\textbf{T}} \bar{\textbf{D}}_{i} \bar{\textbf{T}}) \left(\frac{\bar{m}^{2}_{i}}{S_{i}} \left(\frac{K_{i}}{K}\right)^{2} d_{i}^{4} \text{tr }(\bar{\textbf{S}}_{i} \textbf{W}_{i}^{2} \textbf{S}_{i})  \right), \\
&[\textbf{v}(\bar{\textbf{D}}, \textbf{L})]_{g}=\frac{1}{S_{g}} \text{tr }(\bar{\textbf{D}}_{g}\bar{\textbf{T}} \textbf{L} \bar{\textbf{T}}),
\end{align}
for $g,i=1, \dots, G$. The proof of Lemma 3 can be found in Appendix E.

Using Lemma 3, we have the deterministic equivalents of $T_{4}$ and $T_{5}$ (denoted as $\bar{\beta}^{1}_{g,i}(\textbf{A},\textbf{B})$ and $\bar{\beta}^{2}_{g,i}(\textbf{A},\textbf{B}$) in Theorem 2) and hence $\bar{\beta}_{g,i}(\textbf{A},\textbf{B})$. This yields the expression of $\mathbb{E}[Z_{1}]$ as,
\begin{align}
\label{Z_1}
&\mathbb{E}[Z_{1}]=\frac{1}{K} \sum_{i=1}^{G}\sum_{l=1}^{K_{i}} p_{l,i} \frac{\bar{\beta}_{g,i}(\textbf{A},\textbf{B})}{1+m_{i}}  + o(1).
\end{align}

We now study $Z_{3}$. Using matrix inversion lemma and common inverses of resolvents we have,
\begin{align}
&Z_{3}=- \frac{1}{K^3} \sum_{i=1}^{G} \sum_{l=1}^{K_{i}}  \frac{\text{tr } \textbf{A} \textbf{Z}_{g} \textbf{Z}_{g}^{H} \textbf{B} \textbf{T}  \hat{\textbf{h}}_{l,i} \hat{\textbf{h}}_{l,i}^{H} \hat{\textbf{C}}_{[l,i]}^{-1} \hat{\textbf{H}}^{H} \textbf{P} \hat{\textbf{H}} \hat{\textbf{C}}_{[l,i]}^{-1}}{1+\frac{1}{K} \hat{\textbf{h}}_{l,i}^{H} \hat{\textbf{C}}^{-1}_{[l,i]} \hat{\textbf{h}}_{l,i}} \nonumber \\
&+\frac{1}{K^4} \sum_{i=1}^{G} \sum_{l=1}^{K_{i}} \frac{\text{tr } \textbf{A} \textbf{Z}_{g} \textbf{Z}_{g}^{H} \textbf{B} \textbf{T}  \hat{\textbf{h}}_{l,i} \hat{\textbf{h}}_{l,i}^{H} \hat{\textbf{C}}_{[l,i]}^{-1} \hat{\textbf{H}}^{H} \textbf{P} \hat{\textbf{H}} \hat{\textbf{C}}_{[l,i]}^{-1} \hat{\textbf{h}}_{l,i} \hat{\textbf{h}}_{l,i}^{H} \hat{\textbf{C}}_{[l,i]}^{-1}}{(1+\frac{1}{K} \hat{\textbf{h}}_{l,i}^{H} \hat{\textbf{C}}^{-1}_{[l,i]} \hat{\textbf{h}}_{l,i})^2}= Z_{31}+Z_{32}.
\end{align}
We sequentially deal with terms $Z_{31}$ and $Z_{32}$. Using Lemma 2 and some manipulation we have,
\begin{align}
&\mathbb{E}[Z_{31}]=- \frac{1}{K^3} \sum_{i=1}^{G} \sum_{l=1}^{K_{i}} \frac{ \mathbb{E}\left[\hat{\textbf{h}}_{l,i}^{H} \hat{\textbf{C}}_{[l,i]}^{-1} \hat{\textbf{H}}_{[l,i]}^{H} \textbf{P}_{[l,i]} \hat{\textbf{H}}_{[l,i]} \hat{\textbf{C}}_{[l,i]}^{-1} \textbf{A} \textbf{Z}_{g} \textbf{Z}_{g}^{H} \textbf{B} \textbf{T}  \hat{\textbf{h}}_{l,i}\right] }{1+m_{i}} \\
&- \frac{1}{K^3} \sum_{i=1}^{G} \sum_{l=1}^{K_{i}} p_{l,i}\frac{\mathbb{E}\left[ \hat{\textbf{h}}_{l,i}^{H} \hat{\textbf{C}}_{[l,i]}^{-1} \hat{\textbf{h}}_{l,i} \hat{\textbf{h}}_{l,i}^{H} \hat{\textbf{C}}_{[l,i]}^{-1} \textbf{A} \textbf{Z}_{g} \textbf{Z}_{g}^{H} \textbf{B} \textbf{T}  \hat{\textbf{h}}_{l,i} \right]}{1+m_{i}} +o(1)=\Psi_{1}+\Psi_{2}.
\end{align}
\normalsize
Assuming $\textbf{Z}_{g}$ to be deterministic, we can use Lemmas 6, 3 and 5  from \cite{linearprecoding1} and Corollary 1 from \cite{theorem} sequentially to obtain,
\begin{align}
\label{Psi21}
&\Psi_{2}=- \frac{1}{K} \sum_{i=1}^{G} \sum_{l=1}^{K_{i}} p_{l,i}\frac{ \mathbb{E}\left[\frac{1}{K_{i}} \text{tr } \left(\tilde{\boldsymbol{\Phi}}_{i} \hat{\textbf{C}}^{-1} \right) \right] \mathbb{E}[\frac{1}{K_{i}} \text{tr }  \left(\tilde{\boldsymbol{\Phi}}_{i}  \textbf{T} \textbf{A} \textbf{Z}_{g} \textbf{Z}_{g}^{H} \textbf{B} \textbf{T} \right)] }{1+m_{i}} + o(1).
\end{align}
Extending the analysis to random $\textbf{Z}_{g}$ based on the Fubini Theorem and using Lemma 2 we have,
\begin{align}
\label{Psi2}
&\Psi_{2}=- \frac{1}{K} \sum_{i=1}^{G} \sum_{l=1}^{K_{i}} p_{l,i}\frac{m_{i} \bar{\beta}_{g,i}(\textbf{A},\textbf{B})}{1+m_{i}} + o(1).
\end{align}

The term $\Psi_{1}$ is compensated by $Z_{2}$. To see this, observe that the first order of the term does not change if we substitute $\hat{\textbf{H}}_{[l,i]}$ by $\hat{\textbf{H}}$ and $\textbf{P}_{[l,i]}$ by \textbf{P} and then apply trace lemma, rank-one perturbation lemma (Lemmas 3 and 5 from \cite{linearprecoding1}) and Fubini theorem. 

Finally, we deal with $Z_{32}$. Using Lemma 2 and some manipulation we have,
\begin{align}
&\mathbb{E}[Z_{32}]=\frac{1}{K^4} \sum_{i=1}^{G} \sum_{l=1}^{K_{i}} \Big(\frac{ \mathbb{E}\left[\hat{\textbf{h}}_{l,i}^{H} \hat{\textbf{C}}_{[l,i]}^{-1} \textbf{A} \textbf{Z}_{g} \textbf{Z}_{g}^{H} \textbf{B} \textbf{T}  \hat{\textbf{h}}_{l,i} \hat{\textbf{h}}_{l,i}^{H} \hat{\textbf{C}}_{[l,i]}^{-1} \hat{\textbf{H}}_{[l,i]}^{H} \textbf{P}_{[l,i]} \hat{\textbf{H}}_{[l,i]} \hat{\textbf{C}}_{[l,i]}^{-1} \hat{\textbf{h}}_{l,i} \right] }{(1+m_{i})^2} \nonumber \\
&+ p_{l,i} \frac{\mathbb{E}\left[\hat{\textbf{h}}_{l,i}^{H} \hat{\textbf{C}}_{[l,i]}^{-1} \textbf{A} \textbf{Z}_{g} \textbf{Z}_{g}^{H} \textbf{B} \textbf{T}  \hat{\textbf{h}}_{l,i} \hat{\textbf{h}}_{l,i}^{H} \hat{\textbf{C}}_{[l,i]}^{-1} \hat{\textbf{h}}_{l,i} \hat{\textbf{h}}_{l,i}^{H} \hat{\textbf{C}}_{[l,i]}^{-1} \hat{\textbf{h}}_{l,i} \right]}{(1+m_{i})^2} \Big) + o(1). \nonumber
\end{align}
Analogously to before, $\mathbb{E}[Z_{32}]$ can be simplified as,
\begin{align}
&\mathbb{E}[Z_{32}]=\frac{1}{K} \sum_{i=1}^{G} \sum_{l=1}^{K_{i}} \frac{\mathbb{E}\left[\frac{1}{K_{i}} \text{tr }  \left(\tilde{\boldsymbol{\Phi}}_{i}  \hat{\textbf{C}}^{-1} \textbf{A} \textbf{Z}_{g} \textbf{Z}_{g}^{H} \textbf{B} \textbf{T} \right)\right] \mathbb{E}\left[\frac{1}{K^2} \text{tr } \left( \boldsymbol{\Phi}_{i} \hat{\textbf{C}}^{-1} \hat{\textbf{H}}^{H} \textbf{P} \hat{\textbf{H}} \hat{\textbf{C}}^{-1}  \right)\right]}{(1+m_{i})^2}  \nonumber \\
&+\frac{1}{K} \sum_{i=1}^{G} \sum_{l=1}^{K_{i}} p_{l,i} \frac{\mathbb{E}\left[\frac{1}{K_{i}} \text{tr }  \left(\tilde{\boldsymbol{\Phi}}_{i}  \hat{\textbf{C}}^{-1} \textbf{A} \textbf{Z}_{g} \textbf{Z}_{g}^{H} \textbf{B} \textbf{T} \right) \right] \left(\mathbb{E}\left[\frac{1}{K_{i}} \text{tr } \left(\tilde{\boldsymbol{\Phi}}_{i} \hat{\textbf{C}}^{-1} \right) \right]\right)^2 }{(1+m_{i})^2} + o(1),\nonumber 
\end{align}
\begin{align}
\label{Z32}
&=\frac{1}{K} \sum_{i=1}^{G} \sum_{l=1}^{K_{i}} \frac{\bar{\beta}_{g,i}(\textbf{A},\textbf{B}) \chi^{o}_{i}}{(1+m_{i})^2}  +\frac{1}{K} \sum_{i=1}^{G} \sum_{l=1}^{K_{i}} p_{l,i}  \frac{m_{i}^2 \bar{\beta}_{g,i}(\textbf{A},\textbf{B})}{(1+m_{i})^2} + o(1),
\end{align}
where $\chi^{o}_{i}$ is defined in (\ref{chi_11}). Combining (\ref{Z_1}), (\ref{Psi2}) and (\ref{Z32}), we obtain,
\begin{align}
\label{chi_bar}
&\kappa^{o}_{g}(\textbf{A},\textbf{B}) = \frac{1}{K} \sum_{i=1}^{G} \sum_{l=1}^{K_{i}} \frac{\bar{\beta}_{g,i}(\textbf{A},\textbf{B})}{(1+m_{i})^2} (p_{l,i}+\chi^{o}_{i}) +o(1).
\end{align}

\subsection{ Deterministic equivalents of $\xi(\textbf{A},\textbf{B})$ and $\chi_{g}$:}

Using similar argument for $\xi(\textbf{A},\textbf{B})=\frac{1}{K^2} \text{tr } \textbf{A} \textbf{B} \hat{\textbf{C}}^{-1} \hat{\textbf{H}}^{H} \textbf{P} \hat{\textbf{H}} \hat{\textbf{C}}^{-1}$ as done for $\kappa_{g}(\textbf{A},\textbf{B})$ in (\ref{control}), we have $\xi(\textbf{A},\textbf{B})- \xi^{o}(\textbf{A},\textbf{B}) \xrightarrow[N \rightarrow \infty]{a.s.} 0$, where $\xi^{o}(\textbf{A},\textbf{B})=\mathbb{E}[\xi(\textbf{A},\textbf{B})]$. 

Repeating the same steps as done for $\kappa^{o}_{g}(\textbf{A},\textbf{B})$, we obtain,
\begin{align}
\label{chi_tilde}
&\xi^{o}(\textbf{A},\textbf{B}) = \frac{1}{K} \sum_{i=1}^{G} \sum_{l=1}^{K_{i}} \frac{\tilde{\beta}_{i}(\textbf{A},\textbf{B})}{(1+m_{i})^2} (p_{l,i}+\chi^{o}_{i}) + o(1),
\end{align}
where $\tilde{\beta}_{i}(\textbf{A},\textbf{B})=\frac{d_{i}^2}{K} \left(\sum_{j=1}^{S_{i}} \frac{\bar{s}_{i,j} u'_{i}(\bar{\textbf{D}}_{i}, \textbf{A}\textbf{B})}{(1+\frac{K_{i}}{K} d_{i}^2 \bar{m}_{i} \bar{s}_{i,j} \delta_{i}(\bar{\textbf{D}}_{i}))^2} + \frac{S_{i}}{\rho_{tr}} u'_{i}(\tilde{\textbf{D}}_{i}, \textbf{A}\textbf{B}) \right)$, where $u'_{i}(\textbf{R}_{i},\textbf{L})$ has been defined in Lemma 3.

Plugging (\ref{chi_bar}) and (\ref{chi_tilde}) into (\ref{chi_11}) yields the deterministic equivalent of $\chi^{o}_{g}$ as,
\begin{align}
\label{chi_fin}
&\chi^{o}_{g}=\frac{d_{g}^2 }{K} \sum_{i=1}^{G} \sum_{l=1}^{K_{i}} (p_{l,i}+\chi^{o}_{i}) \left( \frac{\bar{\beta}_{g,i}(\textbf{R}_{BS_{g}}\textbf{Q}_{g},\textbf{Q}_{g}^{H}\textbf{R}_{BS}^{H})}{(1+m_{i})^2} + \frac{1}{\rho_{tr}} \frac{\tilde{\beta}_{i}(\textbf{R}_{BS_{g}}\textbf{Q}_{g},\textbf{Q}_{g}^{H}\textbf{R}_{BS}^{H})}{(1+m_{i})^2} \right).
\end{align}
Now $\boldsymbol{\chi}^{o}=[\chi^{o}_{1}, \chi^{o}_{2}, \dots, \chi^{o}_{g}]^{T}$ can be expressed as a system of linear equations as,
\begin{align}
&\boldsymbol{\chi}^{o} = (\textbf{I}_{N}-\bar{\textbf{J}})^{-1} \bar{\textbf{v}}, \\
&\text{where, } [\bar{\textbf{J}}]_{g,i}=d_{g}^{2} \frac{K_{i}}{K} \frac{1}{(1+m_{i})^2} \left( \bar{\beta}_{g,i} (\textbf{R}_{BS_{g}}\textbf{Q}_{g},\textbf{Q}_{g}^{H}\textbf{R}_{BS}^{H}) + \frac{1}{\rho_{tr}} \tilde{\beta}_{i}(\textbf{R}_{BS_{g}}\textbf{Q}_{g},\textbf{Q}_{g}^{H}\textbf{R}_{BS}^{H}) \right), \\
&\bar{\textbf{v}}_{g} =d_{g}^{2} \frac{1}{K} \sum_{i=1}^{G} \sum_{l=1}^{K_{i}} \frac{p_{l,i}}{(1+m_{i})^2} \left( \bar{\beta}_{g,i} (\textbf{R}_{BS_{g}}\textbf{Q}_{g},\textbf{Q}_{g}^{H}\textbf{R}_{BS}^{H}) + \frac{1}{\rho_{tr}} \tilde{\beta}_{i}(\textbf{R}_{BS_{g}}\textbf{Q}_{g},\textbf{Q}_{g}^{H}\textbf{R}_{BS}^{H}) \right).
\end{align}

Also, plugging (\ref{chi_fin}) into (\ref{chi_bar}) and (\ref{chi_tilde}) completes the deterministic equivalents of $\kappa^{o}_{g}(\textbf{A},\textbf{B})$ and $\xi^{o}(\textbf{A},\textbf{B})$ respectively. These terms will later be needed in the proof of Theorem 3.

\section{Proof of Lemma 3}

To derive the deterministic equivalent of $u'_{g}(\textbf{R}_{g}, \textbf{L})=\mathbb{E}[\frac{1}{S_{g}}\text{tr } \textbf{R}_{g} \textbf{T} \textbf{L} \textbf{T}]$, where $\textbf{R}_{g}$ and $\textbf{L}$, $g= 1, \dots, G$ are deterministic matrices,  we use the results from Appendix B in (\ref{delta_ex}) and (\ref{Th1_1}) along with $||f_{g}(\textbf{R}_{g})-\delta_{g}(\textbf{R}_{g})||\xrightarrow[N \rightarrow \infty]{a.s.}$ to have $\frac{1}{S_{g}} \text{tr } \textbf{R}_{g} \textbf{T} - \delta_{g}(\textbf{R}_{g}) \xrightarrow[N \rightarrow \infty]{a.s.} 0$, where $\textbf{T}=(\sum_{i=1}^{G} \bar{m}_{i} \tilde{\boldsymbol{\Phi}}_{i} +\alpha \textbf{I}_{N})^{-1}$ and $\delta_{g}(\textbf{R}_{g})=\frac{1}{S_{g}}\textbf{R}_{g} \bar{\textbf{T}}$, where $\bar{\textbf{T}}$ is given by (\ref{T_bar}). To this end  note that,
\begin{align}
\label{label22}
&u'_{g}(\textbf{R}_{g}, \textbf{L})= \mathbb{E}\left[\frac{d}{dl} \frac{1}{S_{g}}\text{tr } \textbf{R}_{g} (\textbf{T}^{-1}- l \textbf{L})^{-1}|_{l=0}\right].
\end{align}
The deterministic equivalent of $\frac{1}{S_{g}}\text{tr } \textbf{R}_{g} (\textbf{T}^{-1}- l \textbf{L})^{-1}$ is obtained using Corollary 1 from \cite{theorem} as,
\begin{align}
\label{label23}
\frac{1}{S_{g}} \text{tr }\textbf{R}_{g} \left(\sum_{z=1}^{G} \bar{m}_{z} \tilde{\boldsymbol{\Phi}}_{z} -l \textbf{L} + \alpha \textbf{I}_{N} \right)^{-1} - \frac{1}{S_{g}} \text{tr }\textbf{R}_{g} (\sum_{z=1}^{G} \bar{F}_{z} \bar{\textbf{D}}_{z} +\frac{K_{z}}{K} d_{z}^2 \frac{\bar{m}_{z}}{\rho_{tr}} \tilde{\textbf{D}}_{z} &-l \textbf{L} +\alpha \textbf{I}_{N})^{-1}  \nonumber \\
& \xrightarrow[N \rightarrow \infty]{a.s.} 0,
\end{align}
where $F_{g}({\textbf{R}}_{g},\textbf{L}), \bar{F}_{g}$ are defined as the unique solution to,
\begin{align}
&\bar{F}_{g}=\frac{1}{S_{g}} \sum_{j=1}^{S_{g}} \frac{\frac{K_{g}}{K} d_{g}^2 \bar{m}_{g}\bar{s}_{g,j}}{1+\frac{K_{g}}{K} d_{g}^2 F_{g}(\bar{\textbf{D}}_{g},\textbf{L}) \bar{m}_{g} \bar{s}_{g,j}}, \\
&F_{g}({\textbf{R}}_{g},\textbf{L})=\frac{1}{S_{g}} \text{tr } \textbf{R}_{g}(\sum_{z=1}^{G} \bar{F}_{z} \bar{\textbf{D}}_{z} +\frac{K_{z}}{K} d_{z}^2 \frac{\bar{m}_{z}}{\rho_{tr}} \tilde{\textbf{D}}_{z}-l \textbf{L}  +\alpha \textbf{I}_{N})^{-1},
\end{align}
such that $(F_{g}({\textbf{R}}_{g},\textbf{L}),\bar{F}_{g}) \geq 0$. Substituting $\bar{F}_{g}$ in $F_{g}({\textbf{R}}_{g},\textbf{L})$, we have,
\begin{align}
\label{F_exp}
&F_{g}({\textbf{R}}_{g},\textbf{L})=\frac{1}{S_{g}} \text{tr } \textbf{R}_{g}\left(\sum_{z=1}^{G} \frac{1}{S_{z}} \sum_{j=1}^{S_{z}} \frac{\frac{K_{z}}{K} d_{z}^2 \bar{m}_{z}\bar{s}_{z,j}}{1+\frac{K_{z}}{K} d_{z}^2 F_{z}(\bar{\textbf{D}}_{z},\textbf{L}) \bar{m}_{z} \bar{s}_{z,j}} \bar{\textbf{D}}_{z} +\frac{K_{z}}{K} d_{z}^2 \frac{\bar{m}_{z}}{\rho_{tr}} \tilde{\textbf{D}}_{z}-l \textbf{L}  +\alpha \textbf{I}_{N}\right)^{-1}
\end{align}
Note that $F_{g}({\textbf{R}}_{g},\textbf{L})$ reduces to $\delta_{g}(\textbf{R}_{g})$ for $l=0$.

The expression of $u'_{g}({\textbf{R}}_{g}, \textbf{L})$ can be obtained by using (\ref{label22}) and (\ref{label23}) to get $u'_{g}({\textbf{R}}_{g}, \textbf{L}) = \frac{d}{dl} F_{g}({\textbf{R}}_{g},\textbf{L})|_{l=0} + o(1)$ resulting in,
\begin{align}
\label{M_der_app}
&u'_{g}(\textbf{R}_{g}, \textbf{L})=\frac{1}{S_{g}} \text{tr } \textbf{R}_{g} \left[\bar{\textbf{T}} \left(\sum_{z=1}^{G} \frac{\bar{\textbf{D}}_{z} \bar{m}_{z}^2 }{S_{z}} \left(\frac{K_{z}}{K}\right)^2 d_{z}^4  \text{tr } (\bar{\textbf{S}}_{z} \textbf{W}_{z}^2 \bar{\textbf{S}}_{z}) u'_{z}(\bar{\textbf{D}}_{z}, \textbf{L})  + \textbf{L} \right) \bar{\textbf{T}}  \right] +o(1),
\end{align}
where $\textbf{W}_{i}=\left( \textbf{I}_{S_{i}} + \frac{K_{i}}{K} d_{i}^2 \bar{m}_{i} \delta_{i}(\bar{\textbf{D}}_{i}) \bar{\textbf{S}}_{i} \right)^{-1}$.

Note that $u'_{g}(\textbf{R}_{g}, \textbf{L})$ depends on the values of $u'_{z}(\bar{\textbf{D}}_{z}, \textbf{L})$. The latter can be expressed as a system of linear equations by solving (\ref{M_der_app}) for $\textbf{R}_{g}=\bar{\textbf{D}}_{g}$. This system is represented by (\ref{M_der_eq1}).

\section{Proof of Theorem 3}

The deterministic equivalents of the energy term $|\mathbb{E}[\textbf{h}_{k,g}^{H} \hat{\textbf{V}} \hat{\textbf{h}}_{k,g}]|^{2}$, the term $\Theta$ of the power normalization, the interference term $\mathbb{E}[\textbf{h}_{k,g}^{H} \hat{\textbf{V}} \hat{\textbf{H}}_{[k,g]}^{H} \textbf{P}_{[k,g]} \hat{\textbf{H}}_{[k,g]} \hat{\textbf{V}} \textbf{h}_{k,g}]$ and the variance term \\ $\text{var }(\textbf{h}_{k,g}^{H} \hat{\textbf{V}} \hat{\textbf{h}}_{k,g}) $ are worked out separately to yield the deterministic equivalent of the SINR. 

\vspace{-.15in}
\subsection{ Deterministic equivalent of $|\mathbb{E}[\textbf{h}_{k,g}^{H} \hat{\textbf{V}} \hat{\textbf{h}}_{k,g}]|^{2}$:}
Note that $\textbf{h}_{k,g}^{H} \hat{\textbf{V}} \hat{\textbf{h}}_{k,g}$ can be written as $\frac{1}{K} \textbf{h}_{k,g}^{H} \hat{\textbf{C}}^{-1} \hat{\textbf{h}}_{k,g}$, where $\hat{\textbf{C}}=\frac{1}{K} \hat{\textbf{H}}^{H} \hat{\textbf{H}} + \alpha \textbf{I}_{N}$. In order to remove the dependency of $\hat{\textbf{C}}$ on $\hat{\textbf{h}}_{k,g}$, we use common inverses of resolvents lemma (Lemma 2 from \cite{linearprecoding1}) to get,
\vspace{-.15in}
\begin{align}
&\textbf{h}_{k,g}^{H} \hat{\textbf{V}} \hat{\textbf{h}}_{k,g}=\frac{1}{K} \textbf{h}_{k,g}^{H} \hat{\textbf{C}}^{-1}_{[k,g]} \hat{\textbf{h}}_{k,g} - \frac{1}{K^{2}}  \frac{\textbf{h}_{k,g}^{H} \hat{\textbf{C}}^{-1}_{[k,g]} \hat{\textbf{h}}_{k,g}  \hat{\textbf{h}}_{k,g}^{H} \hat{\textbf{C}}^{-1}_{[k,g]} \hat{\textbf{h}}_{k,g}}{1+\frac{1}{K} \hat{\textbf{h}}_{k,g}^{H} \hat{\textbf{C}}^{-1}_{[k,g]} \hat{\textbf{h}}_{k,g} }.
\end{align}
The deterministic equivalents of both $\frac{1}{K} \hat{\textbf{h}}_{k,g}^{H} \hat{\textbf{C}}^{-1}_{[k,g]} \hat{\textbf{h}}_{k,g}$ and $\frac{1}{K} {\textbf{h}}_{k,g}^{H} \hat{\textbf{C}}^{-1}_{[k,g]} \hat{\textbf{h}}_{k,g}$ have been derived in Lemma 2 as $m_{g}$ and $h_{g}$ respectively. Note that Lemma 2 not only implies almost sure convergence but also convergence in mean. Therefore, by the dominated convergence theorem and the continuous mapping theorem, we have
\begin{align}
&|\mathbb{E}[\textbf{h}_{k,g}^{H} \hat{\textbf{V}} \hat{\textbf{h}}_{k,g}]|^{2}- \frac{h^2_{g}}{(1+m_{g})^2} \xrightarrow[N \rightarrow \infty]{a.s.} 0.
\end{align}

\vspace{-.1in}
\subsection{Deterministic equivalent of $\Theta=\mathbb{E}[\text{tr } \textbf{P} \hat{\textbf{H}} \hat{\textbf{V}}^{2} \hat{\textbf{H}}^{H}]$: }
\vspace{-.35in}
\begin{align}
&\Theta= \mathbb{E}\left[\frac{1}{K^2} \text{tr } \textbf{P} \hat{\textbf{H}} \hat{\textbf{C}}^{-1} \hat{\textbf{C}}^{-1} \hat{\textbf{H}}^{H}\right] =\mathbb{E}\left[\frac{1}{K^2} \text{tr }  \hat{\textbf{C}}^{-1} \hat{\textbf{H}}^{H}  \textbf{P} \hat{\textbf{H}} \hat{\textbf{C}}^{-1}\right] =\xi^{o}(\textbf{I}_{N},\textbf{I}_{N}),
\end{align}
where $\xi^{o}(\textbf{A},\textbf{B})$ was derived in Appendix D and is given by (\ref{chi_tilde}).

\vspace{-.1in}
\subsection{Deterministic equivalent of $\mathbb{E}[\textbf{h}_{k,g}^{H} \hat{\textbf{V}} \hat{\textbf{H}}_{[k,g]}^{H} \textbf{P}_{[k,g]} \hat{\textbf{H}}_{[k,g]} \hat{\textbf{V}} \textbf{h}_{k,g}]$:} \vspace{-.05in}
Denote $\textbf{h}_{k,g}^{H} \hat{\textbf{V}} \hat{\textbf{H}}_{[k,g]}^{H} \textbf{P}_{[k,g]} \hat{\textbf{H}}_{[k,g]} \hat{\textbf{V}} \textbf{h}_{k,g}$ as $\Upsilon_{k,g}$ and use Lemma 2 from \cite{linearprecoding1} to decompose it as, \vspace{-.05in}
\begin{align}
&\Upsilon_{k,g} =\frac{1}{K^2}\textbf{h}_{k,g}^{H} \hat{\textbf{C}}^{-1}_{[k,g]} \hat{\textbf{H}}_{[k,g]}^{H} \textbf{P}_{[k,g]} \hat{\textbf{H}}_{[k,g]} \hat{\textbf{C}}^{-1}_{[k,g]} \textbf{h}_{k,g}  - \frac{\textbf{h}_{k,g}^{H} \hat{\textbf{C}}^{-1}_{[k,g]} \hat{\textbf{H}}_{[k,g]}^{H} \textbf{P}_{[k,g]} \hat{\textbf{H}}_{[k,g]} \hat{\textbf{C}}^{-1}_{[k,g]} \hat{\textbf{h}}_{k,g} \hat{\textbf{h}}_{k,g}^{H} \hat{\textbf{C}}^{-1}_{[k,g]} \textbf{h}_{k,g}}{K^3(1+\frac{1}{K}\hat{\textbf{h}}_{k,g}^{H} \hat{\textbf{C}}^{-1}_{[k,g]} \hat{\textbf{h}}_{k,g})}\nonumber \\ 
&-\frac{1}{K^3}\frac{\textbf{h}_{k,g}^{H} \hat{\textbf{C}}^{-1}_{[k,g]} \hat{\textbf{h}}_{k,g} \hat{\textbf{h}}_{k,g}^{H} \hat{\textbf{C}}^{-1}_{[k,g]} \hat{\textbf{H}}_{[k,g]}^{H} \textbf{P}_{[k,g]} \hat{\textbf{H}}_{[k,g]} \hat{\textbf{C}}^{-1}_{[k,g]} \textbf{h}_{k,g} }{1+\frac{1}{K}\hat{\textbf{h}}_{k,g}^{H} \hat{\textbf{C}}^{-1}_{[k,g]} \hat{\textbf{h}}_{k,g}} \nonumber \\ 
&+\frac{1}{K^4}\frac{\textbf{h}_{k,g}^{H} \hat{\textbf{C}}^{-1}_{[k,g]} \hat{\textbf{h}}_{k,g} \hat{\textbf{h}}_{k,g}^{H} \hat{\textbf{C}}^{-1}_{[k,g]} \hat{\textbf{H}}_{[k,g]}^{H} \textbf{P}_{[k,g]} \hat{\textbf{H}}_{[k,g]} \hat{\textbf{C}}^{-1}_{[k,g]} \hat{\textbf{h}}_{k,g} \hat{\textbf{h}}_{k,g}^{H} \hat{\textbf{C}}^{-1}_{[k,g]} \textbf{h}_{k,g}}{(1+\frac{1}{K}\hat{\textbf{h}}_{k,g}^{H} \hat{\textbf{C}}^{-1}_{[k,g]} \hat{\textbf{h}}_{k,g})^2}, \nonumber \\ 
&=X_{1}+X_{2}+X_{3}+X_{4}.
\end{align}
\normalsize
Let us begin by treating $X_{1}$. Using trace lemma and results from Appendix D we have,
\begin{align}
&X_{1}-  \frac{1}{K^2} \text{tr } \textbf{Z}_{g} \textbf{Z}_{g}^{H} \hat{\textbf{C}}^{-1}_{[k,g]} \hat{\textbf{H}}_{[k,g]}^{H} \textbf{P}_{[k,g]} \hat{\textbf{H}}_{[k,g]} \hat{\textbf{C}}^{-1}_{[k,g]} \xrightarrow[N \rightarrow \infty]{a.s.} 0. \\
&\frac{1}{K^2} \text{tr } \textbf{Z}_{g} \textbf{Z}_{g}^{H} \hat{\textbf{C}}^{-1}_{[k,g]} \hat{\textbf{H}}_{[k,g]}^{H} \textbf{P}_{[k,g]} \hat{\textbf{H}}_{[k,g]} \hat{\textbf{C}}^{-1}_{[k,g]}= \kappa_{g}(\textbf{I}_{N},\textbf{I}_{N}) + o(1).
\end{align}
Therefore,
\begin{align}
\label{X_1}
&X_{1}-  \kappa^{o}_{g}(\textbf{I}_{N},\textbf{I}_{N}) \xrightarrow[N \rightarrow \infty]{a.s.} 0,
\end{align}
where $\kappa^{o}_{g}(\textbf{A},\textbf{B})$ was derived in the last section and is given by (\ref{chi_bar}).

Next note that,
\begin{align}
&X_{2}=-Y_{2} \frac{\frac{1}{K} \hat{\textbf{h}}_{k,g}^{H} \hat{\textbf{C}}^{-1}_{[k,g]} \textbf{h}_{k,g}}{1+\frac{1}{K}\hat{\textbf{h}}_{k,g}^{H} \hat{\textbf{C}}^{-1}_{[k,g]} \hat{\textbf{h}}_{k,g}}, \\
&Y_{2}=\frac{1}{K^2} \textbf{h}_{k,g}^{H} \hat{\textbf{C}}^{-1}_{[k,g]} \hat{\textbf{H}}_{[k,g]}^{H} \textbf{P}_{[k,g]} \hat{\textbf{H}}_{[k,g]} \hat{\textbf{C}}^{-1}_{[k,g]} \hat{\textbf{h}}_{k,g}.
\end{align}
Using the definition of $\textbf{h}_{k,g}$ and $\hat{\textbf{h}}_{k,g}$ from (\ref{DS1}) and (\ref{MMSE_estimate}) respectively, we have
\begin{align}
&Y_{2}=\frac{\sqrt{S_{g}} d_{g}}{K^2} \tilde{\textbf{w}}_{k,g}^{H} \textbf{Z}_{g}^{H} \hat{\textbf{C}}^{-1}_{[k,g]} \hat{\textbf{H}}_{[k,g]}^{H} \textbf{P}_{[k,g]} \hat{\textbf{H}}_{[k,g]} \hat{\textbf{C}}^{-1}_{[k,g]} \textbf{R}_{BS_{g}} \textbf{Q}_{g} \left(\sqrt{S_{g}} \textbf{Z}_{g} \tilde{\textbf{w}}_{k,g} + \frac{1}{\sqrt{\rho_{tr}}} \textbf{n}^{tr}_{k,g}\right).
\end{align}
Using the trace lemma and the independence of $\tilde{\textbf{w}}_{k,g}$ and $\textbf{n}^{tr}_{k,g}$ we have,
\begin{align}
&Y_{2}-\frac{ d_{g}}{K^2} \text{tr } \textbf{R}_{BS_{g}} \textbf{Q}_{g} \textbf{Z}_{g} \textbf{Z}_{g}^{H} \hat{\textbf{C}}^{-1}_{[k,g]} \hat{\textbf{H}}_{[k,g]}^{H} \textbf{P}_{[k,g]} \hat{\textbf{H}}_{[k,g]} \hat{\textbf{C}}^{-1}_{[k,g]}     \xrightarrow[N \rightarrow \infty]{a.s.} 0.
\end{align}
Note that $\frac{1}{K^2} \text{tr } \textbf{R}_{BS_{g}} \textbf{Q}_{g} \textbf{Z}_{g} \textbf{Z}_{g}^{H} \hat{\textbf{C}}^{-1}_{[k,g]} \hat{\textbf{H}}_{[k,g]}^{H} \textbf{P}_{[k,g]} \hat{\textbf{H}}_{[k,g]} \hat{\textbf{C}}^{-1}_{[k,g]}=\kappa_{g}( \textbf{R}_{BS_{g}} \textbf{Q}_{g} , \textbf{I}_{N}) +o(1)$. Therefore,
\begin{align}
&Y_{2}-  d_{g} \kappa^{o}_{g}( \textbf{R}_{BS_{g}} \textbf{Q}_{g} , \textbf{I}_{N}) \xrightarrow[N \rightarrow \infty]{a.s.} 0,
\end{align}
where $\kappa^{o}_{g}(\textbf{A},\textbf{B}) $ was derived in the last section and is given by (\ref{chi_bar}). Combining this with Lemma 2 yields the deterministic equivalent of $X_{2}$ as,
\begin{align}
\label{X_2}
&X_{2} + \frac{h_{g}}{1+m_{g}} d_{g} \kappa^{o}_{g}( \textbf{R}_{BS_{g}} \textbf{Q}_{g} , \textbf{I}_{N}) \xrightarrow[N \rightarrow \infty]{a.s.} 0.
\end{align}
Similar analysis yields the deterministic equivalent of $X_{3}$ as,
\begin{align}
\label{X_3}
&X_{3} + \frac{h_{g}}{1+m_{g}} d_{g} \kappa^{o}_{g}( \textbf{I}_{N},\textbf{Q}_{g}^{H} \textbf{R}_{BS_{g}}^{H} ) \xrightarrow[N \rightarrow \infty]{a.s.} 0.
\end{align}
Next note that,
\begin{align}
&X_{4}=Y_{4} \frac{\frac{1}{K}\hat{\textbf{h}}_{k,g}^{H} \hat{\textbf{C}}^{-1}_{[k,g]} \textbf{h}_{k,g} \frac{1}{K}{\textbf{h}}_{k,g}^{H} \hat{\textbf{C}}^{-1}_{[k,g]} \hat{\textbf{h}}_{k,g}}{(1+\frac{1}{K}\hat{\textbf{h}}_{k,g}^{H} \hat{\textbf{C}}^{-1}_{[k,g]} \hat{\textbf{h}}_{k,g})^2},
\end{align}
where $Y_{4}=\frac{1}{K^2} \hat{\textbf{h}}_{k,g}^{H} \hat{\textbf{C}}^{-1}_{[k,g]} \hat{\textbf{H}}_{[k,g]}^{H} \textbf{P}_{[k,g]} \hat{\textbf{H}}_{[k,g]} \hat{\textbf{C}}^{-1}_{[k,g]} \hat{\textbf{h}}_{k,g}$. Using Fubini theorem and trace lemma we have $Y_{4}- \frac{1}{K^2} \boldsymbol{\Phi}_{g} \hat{\textbf{C}}^{-1}_{[k,g]} \hat{\textbf{H}}_{[k,g]}^{H} \textbf{P}_{[k,g]} \hat{\textbf{H}}_{[k,g]} \hat{\textbf{C}}^{-1}_{[k,g]}  \xrightarrow[N \rightarrow \infty]{a.s.} 0$. Notice that $\frac{1}{K^2} \boldsymbol{\Phi}_{g} \hat{\textbf{C}}^{-1}_{[k,g]} \hat{\textbf{H}}_{[k,g]}^{H} \textbf{P}_{[k,g]} \hat{\textbf{H}}_{[k,g]} \hat{\textbf{C}}^{-1}_{[k,g]} =\chi_{g}+o(1)$, where $\chi_{g}$ is defined in (\ref{chi_quant}). This yields the deterministic equivalent of $X_{4}$ as,
\begin{align}
\label{X_4}
&X_{4} - \frac{h_{g}^2}{(1+m_{g})^2} \chi_{g}^{o} \xrightarrow[N \rightarrow \infty]{a.s.} 0.
\end{align}
Combining (\ref{X_1}), (\ref{X_2}), (\ref{X_3}) and (\ref{X_4}) yields the deterministic equivalent of $\Upsilon_{k,g}$. The lemmas utilized here not only imply almost sure convergence but also convergence in mean. Therefore,
\begin{align}
&\mathbb{E}[\textbf{h}_{k,g}^{H} \hat{\textbf{V}} \hat{\textbf{H}}_{[k,g]}^{H} \textbf{P}_{[k,g]} \hat{\textbf{H}}_{[k,g]} \hat{\textbf{V}} \textbf{h}_{k,g}]- \Big(\kappa^{o}_{g}(\textbf{I}_{N},\textbf{I}_{N}) - \frac{h_{g}}{1+m_{g}} d_{g} (\kappa^{o}_{g}( \textbf{R}_{BS_{g}} \textbf{Q}_{g}, \textbf{I}_{N}) \nonumber \\
&+ \kappa^{o}_{g}( \textbf{I}_{N},\textbf{Q}_{g}^{H} \textbf{R}_{BS_{g}}^{H}) )+ \frac{h_{g}^2}{(1+m_{g})^2} \chi^{o}_{g} \Big)  \xrightarrow[N \rightarrow \infty]{a.s.} 0.
\end{align}
\vspace{-.3in}
\subsection{Deterministic Equivalent of $\text{var }(\textbf{h}_{k,g}^{H} \hat{\textbf{V}} \hat{\textbf{h}}_{k,g})$}
Define the quantities $a=\textbf{h}_{k,g}^H \hat{\textbf{V}} \hat{\textbf{h}}_{k,g}$ and $\bar{a}=\mathbb{E}[{\textbf{h}}_{k,g}^H \hat{\textbf{V}} \hat{\textbf{h}}_{k,g}]$. By matrix inversion lemma we have $0 \leq a,\bar{a} \leq 1$. Thus $\text{var }(\textbf{h}_{k,g}^{H} \hat{\textbf{V}} \hat{\textbf{h}}_{k,g})=\mathbb{E}[|a-\bar{a}|^2] \leq 2\mathbb{E}[|a-\bar{a}|]$. 
We have already shown that $a-\frac{h_{g}}{1+m_{g}} \xrightarrow[N \rightarrow \infty]{a.s.} 0$. Since $a$ and $\bar{a}$ are bounded, so by dominated convergence theorem we have $\mathbb{E}|a-\bar{a}|\xrightarrow[N \rightarrow \infty]{a.s.} 0$.  Therefore $\text{var }(\textbf{h}_{k,g}^{H} \hat{\textbf{V}} \hat{\textbf{h}}_{k,g})\xrightarrow[N \rightarrow \infty]{a.s.} 0$.

Combining the results of the four subsections completes the proof of Theorem 3.

\vspace{-.15in}
\bibliographystyle{IEEEtran}
\bibliography{bib}

\begin{thebibliography}{10}
\providecommand{\url}[1]{#1}
\csname url@samestyle\endcsname
\providecommand{\newblock}{\relax}
\providecommand{\bibinfo}[2]{#2}
\providecommand{\BIBentrySTDinterwordspacing}{\spaceskip=0pt\relax}
\providecommand{\BIBentryALTinterwordstretchfactor}{4}
\providecommand{\BIBentryALTinterwordspacing}{\spaceskip=\fontdimen2\font plus
\BIBentryALTinterwordstretchfactor\fontdimen3\font minus
  \fontdimen4\font\relax}
\providecommand{\BIBforeignlanguage}[2]{{%
\expandafter\ifx\csname l@#1\endcsname\relax
\typeout{** WARNING: IEEEtran.bst: No hyphenation pattern has been}%
\typeout{** loaded for the language `#1'. Using the pattern for}%
\typeout{** the default language instead.}%
\else
\language=\csname l@#1\endcsname
\fi
#2}}
\providecommand{\BIBdecl}{\relax}
\BIBdecl

\bibitem{massive2}
T.~L. Marzetta, ``Noncooperative cellular wireless with unlimited numbers of
  base station antennas,'' \emph{IEEE Transactions on Wireless Communications},
  vol.~9, no.~11, pp. 3590--3600, November 2010.

\bibitem{massiveMIMOO}
J.~Hoydis, S.~T. Brink, and M.~Debbah, ``Massive {MIMO in the UL/DL} of
  cellular networks: How many antennas do we need?'' \emph{IEEE Journal on
  Selected Areas in Communications}, vol.~31, no.~2, pp. 160--171, Feb. 2013.

\bibitem{massive_Luca}
E.~Bj{\"{o}}rnson, J.~Hoydis, and L.~Sanguinetti, ``Massive {MIMO} has
  unlimited capacity,'' \emph{{IEEE} Transactions on Wireless Communications},
  vol.~17, no.~1, pp. 574--590, 2018.

\bibitem{massive_Luca1}
H.~Yin, D.~Gesbert, M.~Filippou, and Y.~Liu, ``A coordinated approach to
  channel estimation in large-scale multiple-antenna systems,'' \emph{{IEEE}
  Journal on Selected Areas in Communications}, vol.~31, no.~2, pp. 264--273,
  2013.

\bibitem{massive_Luca2}
H.~Huh, G.~Caire, H.~C. Papadopoulos, and S.~A. Ramprashad, ``Achieving
  "massive {MIMO}" spectral efficiency with a not-so-large number of
  antennas,'' \emph{{IEEE} Transactions on Wireless Communications}, vol.~11,
  no.~9, pp. 3226--3239, 2012.

\bibitem{massiveMIMObook1}
T.~L. Marzetta, E.~G. Larsson, H.~Yang, and H.~Q. Ngo, \emph{Fundamentals of
  Massive {MIMO}}.\hskip 1em plus 0.5em minus 0.4em\relax Cambridge, UK:
  Cambridge University Press, 2016.

\bibitem{SINRdeterministic}
S.~Wagner, R.~Couillet, M.~Debbah, and D.~T.~M. Slock, ``Large system analysis
  of linear precoding in correlated {MISO} broadcast channels under limited
  feedback,'' \emph{IEEE Transactions on Information Theory}, vol.~58, no.~7,
  pp. 4509--4537, 2012.

\bibitem{linearprecoding}
A.~Kammoun, A.~Müller, E.~Björnson, and M.~Debbah, ``Linear precoding based
  on polynomial expansion: Large-scale multi-cell {MIMO} systems,'' \emph{IEEE
  Journal of Selected Topics in Signal Processing}, vol.~8, no.~5, pp.
  861--875, Oct. 2014.

\bibitem{massiveMIMObook}
E.~Bj{\"o}rnson, J.~Hoydis, and L.~Sanguinetti, \emph{Massive{ MIMO} Networks:
  Spectral, Energy, and Hardware Efficiency}.\hskip 1em plus 0.5em minus
  0.4em\relax Foundations and Trends in Signal Processing, vol. 11, no. 3-4,
  pp. 154-655, 2017.

\bibitem{gesbert}
D.~Gesbert, H.~B{\"o}lcskei, D.~Gore, and A.~Paulraj, ``Outdoor {MIMO} wireless
  channels: models and performance prediction,'' \emph{IEEE Transactions on
  Communications}, vol.~50, no.~12, pp. 1926--1934, 2002.

\bibitem{gesbert1}
D.~Chizhik, G.~J. Foschini, M.~J. Gans, and R.~A. Valenzuela, ``Keyholes,
  correlations, and capacities of multielement transmit and receive antennas,''
  \emph{IEEE Transactions on Wireless Communications}, vol.~1, no.~2, pp.
  361--368, April 2002.

\bibitem{correlationpresence3}
M.~Chiani, M.~Win, and A.~Zanella, ``On the capacity of spatially correlated
  {MIMO} rayleigh-fading channels,'' \emph{IEEE Transactions on Information
  Theory}, vol.~49, no.~10, pp. 2363--2371, Oct. 2003.

\bibitem{scattering}
R.~R. Muller, ``A random matrix model of communication via antenna arrays,''
  \emph{IEEE Transactions on Information Theory}, vol.~48, no.~9, pp.
  2495--2506, Sep 2002.

\bibitem{doublescattering2}
H.~Shin and M.~Z. Win, ``{MIMO} diversity in the presence of double
  scattering,'' \emph{IEEE Transactions on Information Theory}, vol.~54, no.~7,
  pp. 2976--2996, July 2008.

\bibitem{keyholee2}
\BIBentryALTinterwordspacing
H.~Q. Ngo and E.~G. Larsson, ``No downlink pilots are needed in massive
  {MIMO},'' \emph{CoRR}, vol. abs/1606.02348, 2016. [Online]. Available:
  \url{http://arxiv.org/abs/1606.02348}
\BIBentrySTDinterwordspacing

\bibitem{doublescattering1}
H.~Shin and J.~H. Lee, ``Capacity of multiple-antenna fading channels: spatial
  fading correlation, double scattering, and keyhole,'' \emph{IEEE Transactions
  on Information Theory}, vol.~49, no.~10, pp. 2636--2647, Oct 2003.

\bibitem{doublescattering3}
S.~Yang and J.~C. Belfiore, ``Diversity-multiplexing tradeoff of double
  scattering {MIMO} channels,'' \emph{IEEE Transactions on Information Theory},
  vol.~57, no.~4, pp. 2027--2034, April 2011.

\bibitem{doublescattering4}
S.~Jin, M.~R. McKay, K.~K. Wong, and X.~Gao, ``Transmit beamforming in rayleigh
  product {MIMO} channels: Capacity and performance analysis,'' \emph{IEEE
  Transactions on Signal Processing}, vol.~56, no.~10, pp. 5204--5221, Oct.
  2008.

\bibitem{doublescattering6}
X.~Li, S.~Jin, X.~Gao, and M.~R. McKay, ``Capacity bounds and low complexity
  transceiver design for double-scattering {MIMO} multiple access channels,''
  \emph{IEEE Transactions on Signal Processing}, vol.~58, no.~5, pp.
  2809--2822, May 2010.

\bibitem{levin}
G.~Levin and S.~Loyka, ``Multi-keyhole mimo channels: Asymptotic analysis of
  outage capacity,'' in \emph{2006 IEEE International Symposium on Information
  Theory}, July 2006, pp. 1305--1309.

\bibitem{doublescattering}
J.~Hoydis, R.~Couillet, and M.~Debbah, ``Asymptotic analysis of
  double-scattering channels,'' in \emph{Forty Fifth Asilomar Conference on
  Signals, Systems and Computers, (ACSCC)}, Nov. 2011, pp. 1935--1939.

\bibitem{ourworkRMT1}
Q.-U.-A. Nadeem, A.~Kammoun, M.~Debbah, and M.-S. Alouini, ``Asymptotic
  analysis of regularized zero-forcing in double scattering channels,'' in
  \emph{IEEE Global Communications Conference (GLOBECOM)}, Abu Dhabi, U.A.E.,
  2018.

\bibitem{doublescattering_multicell}
T.~V. Chien, E.~Björnson, and E.~G. Larsson, ``Multi-cell massive{ MIMO}
  performance with double scattering channels,'' in \emph{International
  Workshop on Computer Aided Modelling and Design of Communication Links and
  Networks (CAMAD)}, Oct. 2016, pp. 231--236.

\bibitem{Medard}
M.~Medard, ``The effect upon channel capacity in wireless communications of
  perfect and imperfect knowledge of the channel,'' \emph{IEEE Transactions on
  Information Theory}, vol.~46, no.~3, pp. 933--946, Sep. 2006.

\bibitem{multicellTDD}
J.~Jose, A.~Ashikhmin, T.~L. Marzetta, and S.~Vishwanath, ``Pilot contamination
  and precoding in multi-cell {TDD} systems,'' \emph{IEEE Transactions on
  Wireless Communications}, vol.~10, no.~8, pp. 2640--2651, Aug. 2011.

\bibitem{walid_mutual_information}
W.~Hachem, O.~Khorunzhiy, P.~Loubaton, J.~Najim, and L.~Pastur, ``A new
  approach for mutual information analysis of large dimensional multi-antenna
  channels,'' \emph{IEEE Transactions on Information Theory}, vol.~54, no.~9,
  pp. 3987--4004, 2008.

\bibitem{iterative}
\BIBentryALTinterwordspacing
J.~Hoydis, R.~Couillet, and M.~Debbah, ``Iterative deterministic equivalents
  for the performance analysis of communication systems,'' \emph{CoRR}, vol.
  abs/1112.4167, 2011. [Online]. Available:
  \url{http://arxiv.org/abs/1112.4167}
\BIBentrySTDinterwordspacing

\bibitem{theorem}
R.~Couillet, M.~Debbah, and J.~W. Silverstein, ``A deterministic equivalent for
  the analysis of correlated {MIMO} multiple access channels,'' \emph{IEEE
  Transactions on Information Theory}, vol.~57, no.~6, pp. 3493--3514, June
  2011.

\bibitem{mapping}
P.~Billingsley, \emph{Probability and Measure}.\hskip 1em plus 0.5em minus
  0.4em\relax Hoboken, NJ: Wiley, 1995.

\bibitem{ourworkTCOM}
Q.~Nadeem, A.~Kammoun, M.~Debbah, and M.~Alouini, ``Design of {5G} full
  dimension massive {MIMO} systems,'' \emph{IEEE Transactions on
  Communications}, vol.~66, no.~2, pp. 726--740, Feb 2018.

\bibitem{usergroups}
J.~Nam, A.~Adhikary, J.~Y. Ahn, and G.~Caire, ``Joint spatial division and
  multiplexing: Opportunistic beamforming, user grouping and simplified
  downlink scheduling,'' \emph{IEEE Journal of Selected Topics in Signal
  Processing}, vol.~8, no.~5, pp. 876--890, Oct 2014.

\bibitem{convergence}
A.~Kammoun and \textit{et al.}, ``A central limit theorem for the {SINR at the
  LMMSE} estimator output for large-dimensional signals,'' \emph{IEEE
  Transactions on Information Theory}, vol.~55, no.~11, pp. 5048--5063, Nov.
  2009.

\bibitem{linearprecoding1}
A.~M{\"{u}}ller, A.~Kammoun, E.~Bj{\"{o}}rnson, and M.~Debbah, ``Linear
  precoding based on polynomial expansion: reducing complexity in massive
  {MIMO},'' \emph{{EURASIP} Journal on Wireless Communications and Networking},
  vol. 2016, p.~63, 2016.

\bibitem{Vitali}
E.~C. Titchmarsh, \emph{The Theory of Functions}.\hskip 1em plus 0.5em minus
  0.4em\relax London, UK: Oxford University Press, 1939.

\bibitem{fixedpoint1}
R.~Couillet, J.~Hoydis, and M.~Debbah, ``Random beamforming over quasi-static
  and fading channels: A deterministic equivalent approach,'' \emph{IEEE
  Transactions on Information Theory}, vol.~58, no.~10, pp. 6392--6425, Oct
  2012.

\end{thebibliography}

\end{document}